# Influence of charging conditions on simulated temperature-programmed desorption for hydrogen in metals

A. Díaz[1*], I.I. Cuesta[1], E. Martínez-Pañeda[2], J.M. Alegre[1]

[1]University of Burgos, Escuela Politécnica Superior. Avenida Cantabria s/n, 09006, Burgos, Spain

[2]Imperial College London, Department of Civil and Environmental Engineering. London SW7 2AZ, UK

**Abstract**

Failures attributed to hydrogen embrittlement are a major concern for metals so a better understanding of damage micro-mechanisms and hydrogen diffusion within the metal is needed. Local concentrations depend on transport phenomena including trapping effects, which are usually characterised by a temperature-programmed desorption method often referred to as Thermal Desorption Analysis (TDA). When the hydrogen is released from the specimen during the programmed heating, some desorption peaks are observed that are commonly related to detrapping energies by means of an analytical procedure. The limitations of this approach are revisited here and gaseous hydrogen charging at high temperatures is simulated. This popular procedure enables attaining high concentrations due to the higher solubility of hydrogen at high temperatures. However, the segregation behaviour of hydrogen into traps depends on charging time and temperature. This process and the subsequent cooling alter hydrogen distribution are numerically modelled; it is found that TDA spectra are strongly affected by the charging temperature and the charging time, both for weak and strong traps. However, the influence of ageing time at room temperature after cooling and before desorption is only appreciable for weak traps.

1. **Introduction**

Research on hydrogen embrittlement comprises two coupled phenomena: micromechanisms of hydrogen assisted fracture and characterisation of hydrogen transport processes. Since the first work on this subject was published [1] two empirical facts have been consistently observed: (i) embrittlement increases with exposure time to a hydrogen-containing environment, and (ii) the effects are temporary and disappear after some ageing time in an inert environment. Both factual observations indicate the clear role of diffusion in time-dependent hydrogen embrittlement. However, modelling breakthroughs on hydrogen transport within metals have chronologically gained complexity. Mass balance and the definition of a flux proportional to a concentration gradient, i.e. Fick's laws, constitute the starting point of all diffusion models. Within this framework, some drifting forces might deviate the ideal diffusion behaviour; temperature gradients promote mass transport as the thermophoresis or Soret effect takes place [2]. Additionally, a pressure gradient also modifies hydrogen flux from compression to tensile regions [3]. This latter process is especially worth considering for hydrogen assisted fracture, given the exponential dependence on lattice hydrogen concentration with crack tip hydrostatic stresses [3–5]. In metals, even in the absence of drifting fields as temperature or stress, real transport behaviour regularly deviates from ideal diffusion due to the presence of crystal defects acting as retention sites or traps for hydrogen. Trapping characterisation and modelling is one of the main focus of hydrogen embrittlement research for defects as dislocations, inclusions or grain boundaries play a critical role in plasticity and fracture. Even though hydrogen localisation is challenging, several tests are useful for hydrogen mapping within the metal microstructure at a high or very high spatial resolution [6]. For instance, silver decoration, hydrogen microprint technique (HMT), secondary ion mass spectroscopy (SIMS), scanning Kelvin probe (SKP) and force microscopy SKP (SKPFM), neutron radiography, and atom probe tomography (APT). However, two classical methods of hydrogen measurement are by far the most popular tests for characterising trapping phenomena: electrochemical permeation (EP) and thermal desorption analysis (TDA). Even though this latter method is usually also referred to as TDS (Thermal Desorption Spectroscopy), the term TDA is preferred throughout this work. Both EP and TDA tests are unable to explicitly determine hydrogen segregation in different metal defects since hydrogen desorption is measured over the entire specimen. However, the evolution of exit fluxes can be related to trapping features with the help of numerical approaches. This is where modelling gains its importance: the analysis of hydrogen trapping without spatial resolution requires modifying transport models to realistically reproduce trapping effects. In the present work, TDA modelling is revisited and different conditions are simulated with the aim of improving the characterisation of traps.

The particular objective of the present paper is the study of thermal desorption after gaseous charging. Gaseous charging is associated to lower hydrogen concentrations at room temperature in comparison to electrochemical methods; for this reason, specimens are usually subjected to high pressures (from 10 MPa to 100 MPa [7–11]) and high temperatures (from 50ºC to 600ºC [9–12]) in an $H_2$ environment. Surface damage suffered during electrochemical charging is avoided using gaseous methods so bulk phenomena, embrittlement as well as hydrogen transport, are expected to be better understood without the aggressive electrochemical entry [11]. The equivalence between electrochemical and gaseous charging can be studied through the concept of fugacity [13,14]. Nevertheless, the cooling process within the autoclave and the subsequent transport to TDA equipment might alter the determination of hydrogen concentrations and trapping characterisation. High-temperature charging, which has been overlooked in TDA modelling, and its effects on desorption spectra are here evaluated through finite element simulations.

The analytic approach commonly used to analyse desorption spectra is revisited in Section 2.1. and its limitations discussed. In contrast, a physically-based kinetic approach, which was first presented by McNabb and Foster [15], is described in Section 2.2. The description of the finite element implementation is given in Section 3. The choice of diffusion parameters, aiming at reproducing hydrogen transport in bcc iron, is also discussed. The validity of thermodynamic equilibrium and its equivalence to the general kinetic approach for high vibration frequencies of hydrogen within the metal lattice is demonstrated in Section 4. In addition to frequency effects, the influence of initial concentration is discussed in Section 5 with the aim of assessing a possible bias during trapping energy determination depending on the charging conditions. However, the emphasis is put on gaseous charging modelling and the simulation of subsequent steps: cooling, aging at room temperature and the thermally programmed desorption in which the desorption flux is plotted for the corresponding temperature.

## 2. Modelling approaches
### 2.1. Kissinger's expression

Analytical regression is the most usual mean to find trapping energies, specifically detrapping activation energies $E_d$, as depicted in Figure 1. It is based on the reaction-diffusion equation proposed by Kissinger [16] which, for a temperature $T$, gas constant $R$ and time $t$ reads:

$$\frac{dX}{dt} = A(1-X)^n \exp\left(-\frac{E_d}{RT}\right) \qquad (1)$$

Here, $n$ is the reaction order (usually taken as one) and $X$ is the fraction of trapped hydrogen that has escaped, i.e. $X = (C_{T,0} - C_T)/C_{T,0}$, being $C_{T,0}$ the initial hydrogen concentration in traps and $C_T$ the uniform hydrogen concentration in traps at each moment. Figure 1 clarifies the distinct definition of detrapping activation energy and binding energy $E_b$. The definition of a potential landscape for hydrogen 1D walk is fundamental for diffusion and trapping modelling. Different schemes might be considered but the definition of a two-level system (L: lattice and T: trapping) is a common strategy. More details of potential energy landscapes for hydrogen diffusion might be found in Refs. [17–21].

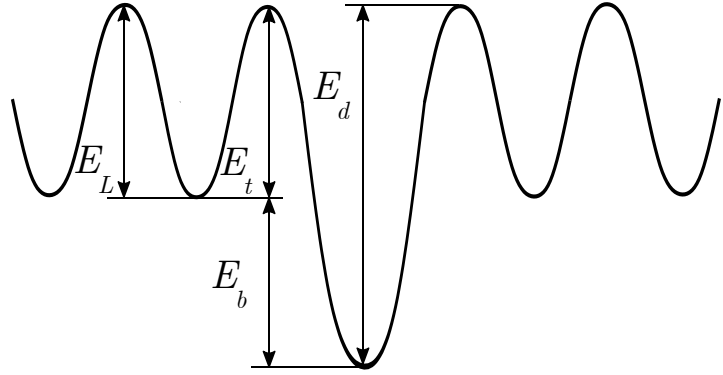

*Figure 1. Schematic definition of trapping, detrapping, binding and lattice energy in a 1D diffusion path.*

Equation (1) assumes that hydrogen concentration in trapping sites is uniform during the entire desorption process, i.e. $C_T(t) = C_T(x,t)$. Analytical regression does not focus on $X$ evolution but on the temperature $T_p$ at which maximum desorption occurs:

$$\frac{d}{dt}\left(\frac{dX}{dt}\right)\bigg|_{T_p} = 0 \tag{2}$$

Kissinger's equation is transferred to TDA assuming temperature as the new dependent variable; a temperature ramp is defined for a constant heating rate $\phi$ such that $T = T_0 + \phi t$. Substituting this ramp in Eq. (1) and differentiating as in (2), the following expression is obtained:

$$\ln\left(\frac{\phi}{T_p^2}\right) = -\frac{E_d}{R}\left(\frac{1}{T_p}\right) + \ln\left(A\frac{R}{E_d}\right) \tag{3}$$

This expression might be linearly fitted plotting $\ln(\phi/T_p^2)$ versus $1/T_p$. Works following this strategy focus on the slope of linear regression to find a detrapping energy. However, as shown in equation (2), $E_d$ also appears in the intersection with the origin with the constant $A$. This constant has been interpreted by Guterl *et al.* [22] and related to vibration frequencies; it also defines a frontier value between detrapping-limited and retrapping-limited regimes.

Assuming this behaviour, detrapping energies are commonly determined by locating desorption peaks at different heating rates. This fitting procedure, first proposed by Kissinger [16], assumes some simplifications and does not take into account the following effects: thickness influence, trapping densities or initial hydrogen concentration. It mus be noted a value of $n = 1$ has been considered, so this reaction order is kept throughout the paper and for every considered regression in Choo-Lee plots. An extensive comparison to the McNabb and Foster's predictions can be found in the work by Wei *et al.* [23]. Kirchheim (2016) [24] derived analytical expressions by solving the associated transport equations for one-level trapping and for multi-trapping thermal desorption. This procedure can be seen as a generalization of Kissinger's equation in which geometry and trap densities also influence the location and shape of TDA spectra.

Various authors have shown that Kissinger's fitting approach underestimates detrapping energies. Legrand *et al.* [25] showed that this underestimation can be profound for thick specimens. In the present work, a cylindrical specimen with a diameter of 10 mm is considered. When trapping effects are taken into account, TDA fitting is a complex inverse problem. Potential solutions are based on Gaussian Deconvolution as a common strategy [25] but future research must focus on advanced regression techniques. In all of these cases, due to the complexity of the equations, analytical solutions are unfeasible and Finite Element modelling is required.

### 2.2. McNabb and Foster's formulation

With the aim of overcoming the limitations exposed in the previous subsection regarding reaction-diffusion regression, a more general model can be established considering a modified mass balance and including a thermodynamic or kinetic relationship between hydrogen concentration in lattice and trapping sites. Neglecting hydrostatic stress-drifted diffusion and thermophoresis, mass balance or Fick's second law might be expressed as [3]:

$$\frac{\partial C_L}{\partial t} + \frac{\partial C_T}{\partial t} = \nabla \cdot \left[ D_{L,0} \exp\left(-\frac{E_L}{RT}\right) \nabla C_L \right] \qquad (4)$$

Where the trapping influence has been explicitly considered by including the term $\partial C_T / \partial t$. Hence, an extra equation must be defined in which a physically based relationship between $C_L$ and $C_T$ is considered. A general kinetic formulation is first presented and then thermodynamic equilibrium is derived as a particular case. It must be highlighted that the jump between neighbour trapping sites is here neglected due to the expected remoteness of traps; a detailed description of generalised transport formulations including multiple traps can be found in the work of Toribio and Kharin (2015) [26].

Kinetic expressions for the variation of hydrogen concentration in traps might be found through a stochastic analysis of the "hops" from traps to lattice sites and vice versa [21]:

$$\frac{\partial C_T}{\partial t} = v_{t,0} \exp\left(-\frac{E_t}{RT}\right)\frac{N_T(1-\theta_T)}{N_L(1-\theta_L)+N_T(1-\theta_T)}C_L \\ -v_{d,0} \exp\left(-\frac{E_d}{RT}\right)\frac{N_L(1-\theta_L)}{N_L(1-\theta_L)+N_T(1-\theta_T)}C_T \quad (5)$$

The first right-term in Eq. (5) represents the trapping process while the second term reproduces detrapping. Here, $\theta_L = C_L/N_L$ and $\theta_T = C_T/N_T$ are the occupancies in lattice and trapping sites, respectively, $N_T$ is the trap density, and $N_L$ is the lattice site density. Frequencies of "hop" attempting from a lattice site to a trap ($v_{t,0}$: trapping process) or from a trap to a lattice site ($v_{d,0}$: detrapping process) are related to the vibration frequency of the hydrogen atom. This expression adds numerical complexity to a solving strategy for the mass balance PDE (4). Usually, Eq. (5) is simplified assuming low occupancy in lattice sites, i.e. $\theta_L <<$ 1 and a trap density much smaller than the density of ideal lattice sites, i.e. $N_T << N_L$, giving equation (6) as a result [15]. While the former assumption is sustained for alloys with low hydrogen solubility, as for bcc iron, the latter condition should be locally checked in regions with a high density of defects, e.g. an extremely deformed region.

$$\frac{\partial C_T}{\partial t} = v_{t,0} \exp\left(-\frac{E_t}{RT}\right)\frac{N_T(1-\theta_T)}{N_L}C_L \\ -v_{d,0} \exp\left(-\frac{E_d}{RT}\right)C_T \quad (6)$$

In their original paper, McNabb and Foster [15] included $N_L$ within a pre-exponential term usually named as $k_0$ corresponding to $v_{t,0}/N_L$. This nomenclature, widely followed, implies different units for $k_0$ (e.g. in m³·mol⁻¹·s⁻¹), and $p_0 = v_{d,0}$ (e.g. in s⁻¹), and dimensional inconsistency, leading in some works to the lack of physical significance for lattice site density or vibration frequencies. For this reason, $v_{t,0}$ and $v_{d,0}$ are used here as pre-exponential coefficients in the McNabb and Foster's formulation. When the number of traps does not change over time, McNabb and Foster's equation is expressed as a function of trapping occupancy:

$$\frac{\partial \theta_T}{\partial t} = v_{t,0} \exp\left(-\frac{E_t}{RT}\right)(1-\theta_T)\theta_L \\ -v_{d,0} \exp\left(-\frac{E_d}{RT}\right)\theta_T \quad (7)$$

When equilibrium is reached, hydrogen "hops" between different types of sites are almost negligible, and the variation in trap occupancy almost zero: $\partial \theta_T/\partial t \approx 0$. Operating in Eq. (7), a relationship between occupancies is found:

$$\frac{\theta_T}{1-\theta_T} = \theta_L \frac{\nu_{t,0}}{\nu_{d,0}} \exp\left(\frac{E_d - E_t}{RT}\right) \qquad (8)$$

This is equivalent to the thermodynamic equilibrium condition, in which the chemical potential of hydrogen in lattice sites is equal to that in traps, as first proposed by Oriani [27]. For $\nu_{t,0} = \nu_{d,0}$, the relationship between occupancies depends only on the equilibrium constant $K_T = \exp(E_b/RT)$, where $E_b$ represents the binding energy defined in Figure 1.

Even though multiple peaks are usually observed experimentally during TDA tests, this does not imply that every peak is associated with a specific defect type. Some other phenomena could result in a desorption peak; hydrogen desorption from a lattice site is observed in cryogenic TDA starting from very low temperatures [28]. It has been also demonstrated that when a non-homogeneous initial hydrogen concentration is imposed, two peaks appear [29]. Two desorption local maximum occurs when two different fronts reach the surface at different times. In the present paper, a one-type trap scenario is considered for the sake of simplicity.

## 3. Finite element model

A 1D axisymmetric model with a radius $a$ is considered with the aim of simulating cylindric specimens that are commonly used for hydrogen charging, TDA and concentration measurements. The partial differential equations that have been presented in the preceding section are implemented in a finite element framework and solved for $\theta_L$ and $\theta_T$ as dependent variables through a backward Euler scheme. A mesh convergence study has been performed, finding that a fine mesh must be used in the exit surface due to the high concentration gradients taking place at the beginning of desorption. A total of 1000 elements are used with a maximum element size of $a/10^4$ in the revolution axis and a minimum size of $a/10^8$ in the exit node.

Desorption at the exit surface is modelled, as in most numerical works, through a Dirichlet boundary condition imposing a zero concentration for lattice sites during the thermally programmed heating. In this 1D axisymmetric approach, only a half of the specimen is modelled so a zero-flux boundary condition is assigned to the node in the symmetry plane ($j_x = -D_L dC_L/dx = 0$ at $x = 0$), whereas at the exit node lattice concentration is fixed ($C_L = 0$ at $x = a$). Zaika *et al.* [30] have considered dynamic boundary conditions for dehydriding modelling. This complex model aims at quantifying the effects of moving phase bounds and volume change during thermal desorption, but it is out of the scope of the present work.

Instantaneous uniform temperature is assumed for all material points; this simplification is only valid for high thermal diffusivities.

Both first-principles calculations [31] and experimental measurements [32] have confirmed that diffusivity is higher in bcc than in fcc iron. The parameters used in the present paper correspond to the results of Ref. [31] considering bcc as ferromagnetic.

Sections 4 and 5 discuss the equilibrium validity and the influence of charging conditions. For this discussion, diffusion parameters related to bcc iron have been considered to investigate the desorption of ferritic steels. The pre-exponential diffusivity, $D_{L,0}$, and the activation energy for lattice diffusion $E_L$ are extracted from [31]. Two assumptions are made: (i) the trapping barrier from a lattice site is equal to the barrier for lattice diffusion, as shown in Figure 1, i.e. $E_L = E_t$; and (ii) the number of lattice sites, $N_L$, corresponds to the preferred tetrahedral sites in bcc iron [21]. The fixed parameters in sections 4 and 5 are thus show in Table 1.

| Pre-exponential diffusivity | $D_{L,0}$ (m$^2$/s) | 1.98×10$^{-7}$ |
|---|---|---|
| Barrier for lattice diffusion | $E_L = E_t$ (kJ/mol) | 8.49 |
| Number of lattice sites | $N_L$ (sites/m$^3$) | 5.095×10$^{29}$ |
| Specimen radius | $a$ (mm) | 5.0 |

*Table 1. Fixed parameters for every TDA simulation.*

The range of simulated binding energies in the following sections lies between 30 and 60 kJ/mol is intended to cover the experimental scatter for different types of defects [33]. The considered density trap ratio $N_T/N_L$ is also very variable depending on the defect type: it can take very low values for stress-free specimens, intermediate values for grain boundaries or very high values, up to 10$^{-3}$, for carbides or highly-deformed locations [34].

### 4. Equilibrium validity

Considering thermodynamic equilibrium between lattice and trapping sites is a common assumption in hydrogen diffusion modelling. McNabb and Foster's equation is equivalent to Oriani's equilibrium when kinetic trapping and detrapping processes occur very fast [27] and/or when the apparent diffusivity is very low [35]. Thus, vibration frequencies are very influential parameters. In the present work, it is assumed that the hydrogen atom vibrates at the same frequency independently of its location site. Also, the hop frequency depends on the trapping and lattice features as well as on energy landscape, as shown in Figure 1 and expressed in the generic kinetic expression (5).

Hydrogen vibration frequency is usually taken to be equal to the Debye frequency ($10^{13}$ s$^{-1}$) [21,36]; calculations based on harmonic transition state theory also give frequency values near the THz range [37] ($10^{12}$ – $10^{13}$ s$^{-1}$). However, Turnbull *et al.* state that the capture and release constants $p_0$ and $k_0$ "are assumed to be fast and only the ratio is considered important" [38]. As previously discussed, this ratio is related to the density of lattice sites; these authors found that $p_0/N_L$ and $k_0$, higher than $10^5$ s$^{-1}$ result in a lack of dependence of desorption peaks or the corresponding peak temperatures. Hurley et al. [29] showed that results converged for frequencies $p_0$ and $k_0$ higher than $10^7$ s$^{-1}$. However, the relationship between oscillation frequencies obtained by harmonic considerations and the pre-exponential constants appearing in McNabb and Foster's equation is still not clear. The limiting value that indicates independence of frequency and equilibrium validity might depend on other factors such as trapping features or hydrogen concentration, so it is evaluated here.

In this section, an initially uniform hydrogen concentration is simulated with $C_{L,0}$ imposed as a numerical initial value, $C_L(x, t = 0) = C_{L,0} = 1.0$ wt ppm, and with the corresponding equilibrium $C_T$. The range in which frequency dependency is valid is investigated firstly considering weak traps with a binding energy of 30 kJ/mol and low-density traps: $N_T = 10^{-6} N_L$. Figure 2.a shows that hydrogen flux rapidly decreases at the beginning of the 800 K/h temperature ramp for the considered low trapping influence and only a small peak is observed at 400 K. A very similar flux drop is also found when strong traps ($E_b$ = 60 kJ/mol) are simulated for the low trap density (Figure 2.b); however, a small peak appears at 700 K for $\nu_0 = 10^3$ s$^{-1}$ and at 1000 K for $\nu_0 = 10^5$ s$^{-1}$. These desorption spectra associated with almost pure lattice diffusion are hardly found in experimental TDA tests. Nevertheless, these simulations demonstrate a small frequency influence only for extremely low $\nu_0$ values.

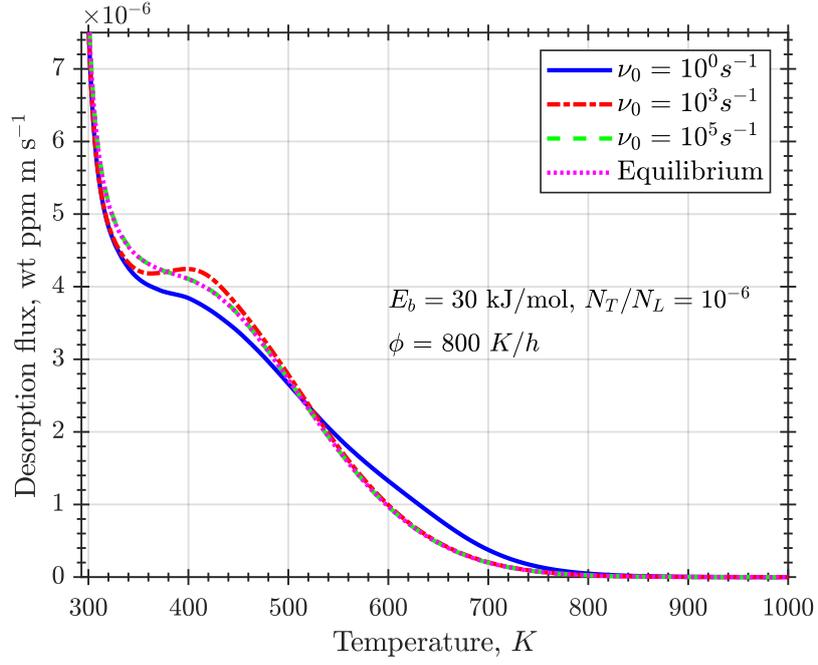

*(a)*

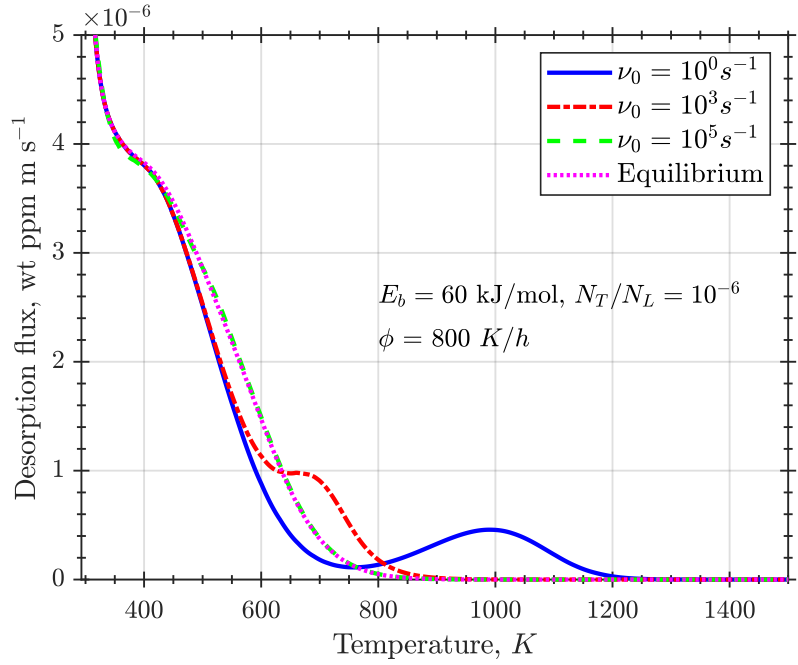

*(b)*

*Figure 2. Influence of vibration frequency on desorption flux during programmed temperature ramp at 800 K/h for $N_T = 10^{-6} N_L$ and (a) weak traps, $E_b = 30$ kJ/mol, or (b) strong traps, $E_b = 60$ kJ/mol.*

Weak traps (30 kJ/mol) can produce desorption peaks at certain temperatures during heating, when a higher defect density is simulated ($N_T = 10^{-3} N_L$) and very low $\nu_0$ values are considered, as shown in Figure 3.a. A decrease in $\nu_0$ promotes a shift to higher temperatures of the detrapping peak. This result is attributed to the slower kinetic exchange and the consequent delay in detrapping process. In this case, the initial flux rapidly increases since the amount of trapped hydrogen is much higher, but it is weakly bound to defect sites. Strong traps (60 kJ/mol) are also simulated for a high trap density and the same heating ramp (800 K/h), resulting in a very similar behaviour but shifted to higher temperatures – see Figure 3.b. For both Figures 3.a and 3.b, a hump-backed curve is obtained for the frequency value of $\nu_0$ = $10^3$ s$^{-1}$. This shoulder appearing only at this specific frequency is attributed to a secondary detrapping process that promotes a second desorption peak. Some authors have stated that peak temperature during thermal desorption depends only on the binding energy independently of the trap density [33,38,39]; comparing Figures 2 and 3, this assertion is confirmed only for some frequencies because the shoulders found in Figure 2, for high frequencies and equilibrium, are not exactly placed at the same temperatures than the results shown in Figure 3.

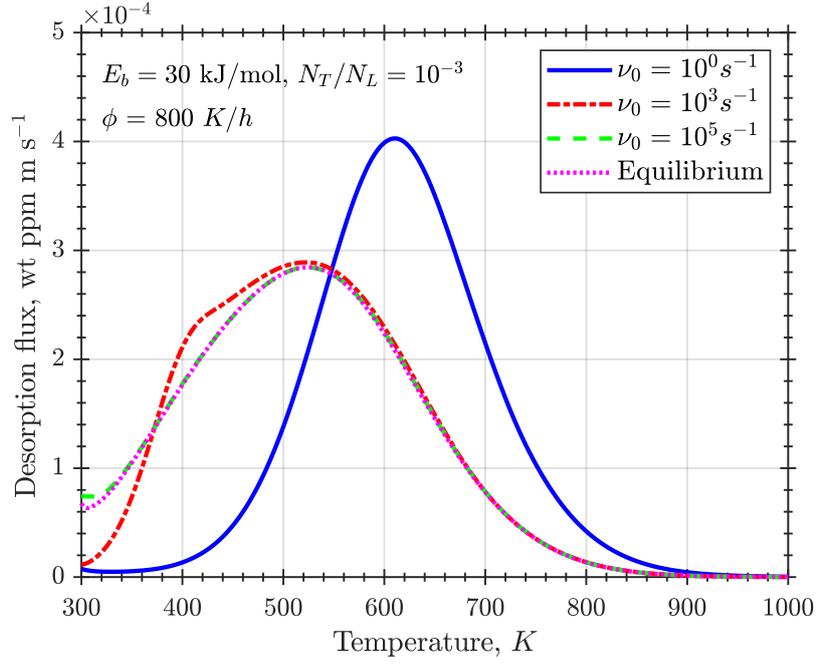

*(a)*

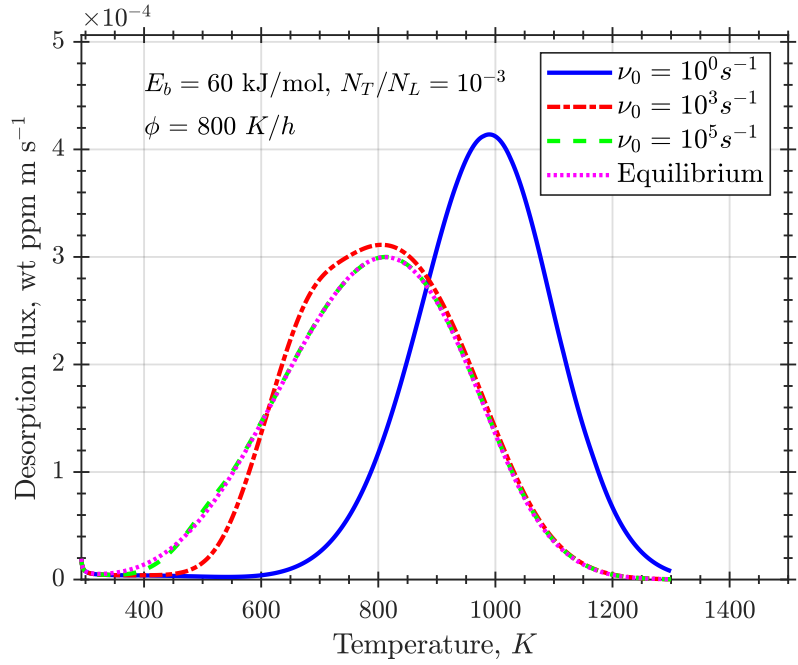

*(b)*

*Figure 3. Influence of vibration frequency on desorption flux during programmed temperature ramp at 800 K/h for $N_T = 10^{-3} N_L$ and (a) weak traps, $E_b$ = 30 kJ/mol, or (b) strong traps, $E_b$ = 60 kJ/mol,*

There appears to be a consensus on the high values of capture and release pre-exponential constants or, equivalently, the hydrogen vibration frequencies; frequencies lower than $10^5$ s$^{-1}$ seem to lack physical significance. Accordingly, the present results demonstrate that Oriani's equilibrium is a valid assumption. We show that, for the fixed parameters here simulated (Table 1), thermodynamic equilibrium can be assumed. The validity of equilibrium has also been assumed by other authors [40] and it has been confirmed for the range of heating rates simulated onwards (400 to 1600 K/h) even though it is not here plotted. Therefore, the influence of trapping features (density $N_T$ and binding energy $E_b$) during thermally programmed desorption at 800 K/h is summarised in Figure 4 considering high vibration frequencies, i.e. equilibrium validity. A rising desorption flux is only observed for high trap densities ($N_T/N_L = 10^{-3}$), whre the stronger traps ($E_b = 60$ kJ/mol) results in a peak shift to higher temperatures. For low trap densities ($N_T/N_L = 10^{-6}$), the flux rapidly decays since the influence of trapping-detrapping processes is small event for high binding energies. It must be noted that these results have been obtained for an input concentration of 1 wt ppm, which represents a typical order of magnitude for hydrogen in bcc iron or other low-solubility hydrogen-metal systems. Nevertheless, considering the equilibrium concentration at traps corresponding to a lattice initial concentration of 1 wt ppm must give huge concentrations that lack physical sense. The initial concentration ranges and numerical considerations are discussed in Section 5.

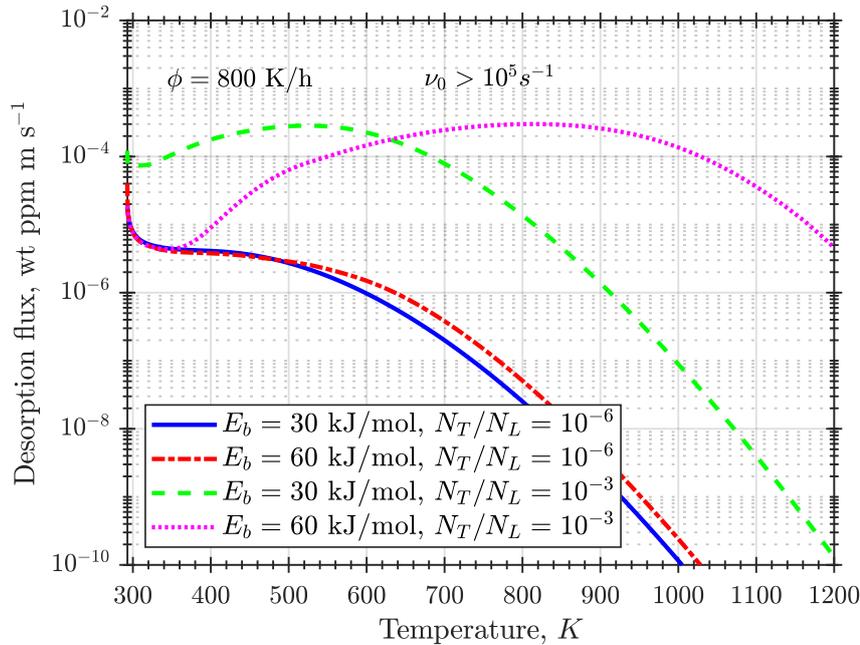

*Figure 4. Influence of trapping parameters on desorption flux during programmed temperature ramp at 800 K/h and high vibration frequencies.*

In order to extend this conclusion to more heating rates; 400, 800, 1200 and 1600 K/h temperature ramps are simulated considering strong traps (60 kJ/mol) and high trap densities ($N_T/N_L$ = 10$^{-3}$). The choice of high trap densities is justified by the results previously in Section 4 so a rising flux and differentiated peaks are obtained. On the other hand, lower binding enerigies but for the same $N_T/N_L$ = 10$^{-3}$ is expected to give similar results but with the peaks shifted towards lower temperatures. Desorption flux versus temperature is plotted in Figure 5 for a low vibration frequency ($\nu_0$ = 1 s$^{-1}$) and four heating rates. The aim is here to assess the deviation of Kissinger's regression from the simulated binding energy and to evaluate whether this deviation is higher at low or high vibration frequencies. For fast heating, the desorption peak is shifted towards higher temperatures whereas the maximum value increases as desorption is occurring in a narrow time interval. It must be noted that the coincidence of curves during the first flux rising, i.e. the left side of the curve, is not observed if desorption flux is plotted against time instead of temperature. Following Kissinger's approach exposed in Section 2.1., a linear regression considering expression (3) is performed taking peak temperatures that have been obtained in FE simulations and their corresponding heating rates. This regression is plotted in Figure 5.b. and the fitted slope ($-E_d/R$) is used to calculate the detrapping energy: $E_d$ = 58.1 kJ/mol. Since the trapping energy has been fixed as 8.49 kJ/mol, a binding energy might be retrieved as: $E_b = E_d - E_t$ = 49.6 kJ/mol, which is 10.4 kJ/mol lower than the input binding energy in the simulation. This fact confirms the expected result that Kissinger's approach underestimates binding energies; however, for the lower frequency, i.e. for a situation far from equilibrium, it is hard to decouple frequency effects and to draw conclusions on the underestimation of energies. For the higher vibration frequency, desorption curves at the same heating rates (400, 800, 1200 and 1600 K/h) are smaller and located at lower temperatures, as shown in Figure 6.b. Linear regression gives a slightly higher detrapping energy ($E_d$ = 58.7 kJ/mol) but still significantly underestimates the imposed binding energy. In order to overcome this underestimation and to improve fitting without the need of numerical simulations, the generalised analytic approach proposed by Kirchheim (2016) [24] can be followed.

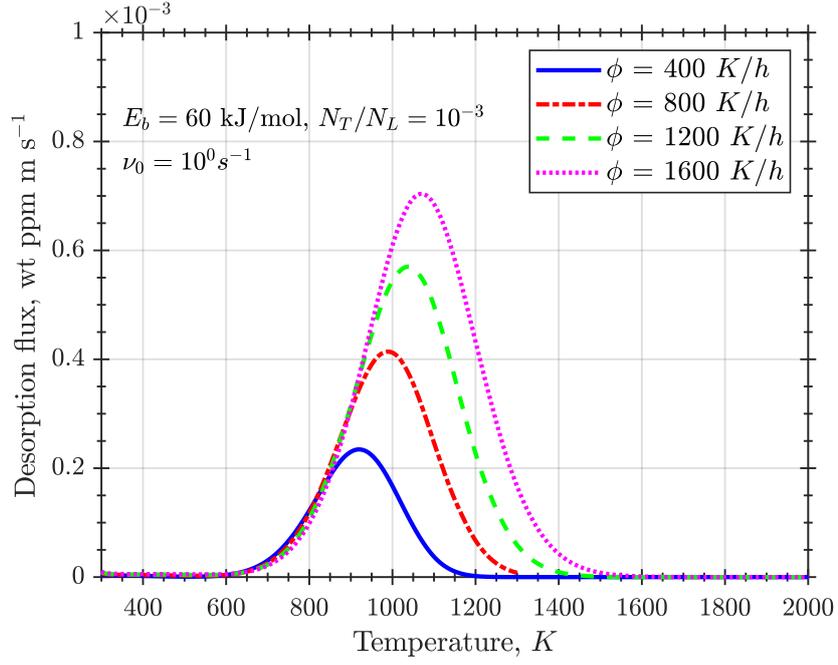

*(a)*

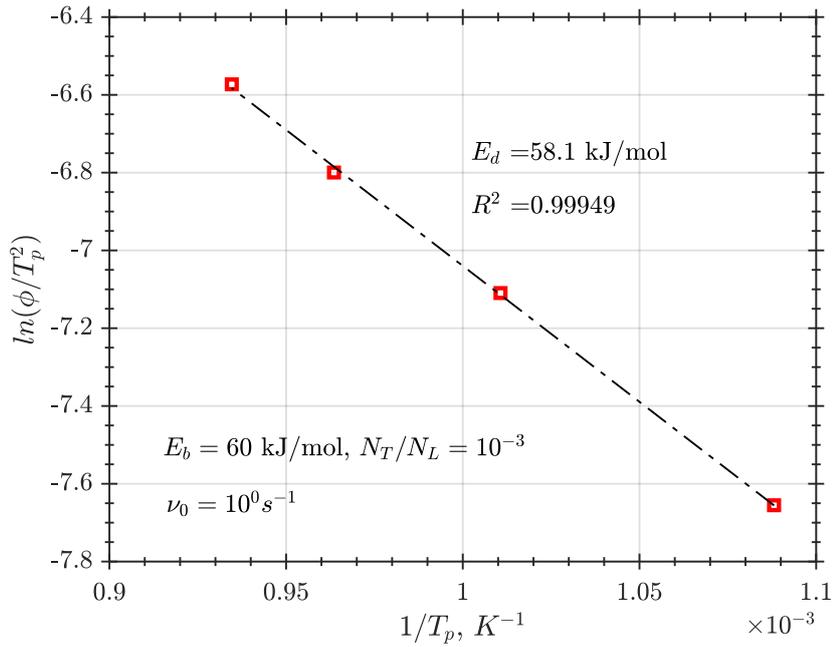

*(b)*

*Figure 5. Results for and low vibration frequency ($\nu_0 = 10^0 \ s^{-1}$), strong traps ($E_b$ = 60 kJ/mol) and high trap densities ($N_T = 10^{-3} N_L$). (a) Desorption flux at different heating rates; (b) detrapping energy determination using peak temperatures at different heating rates.*

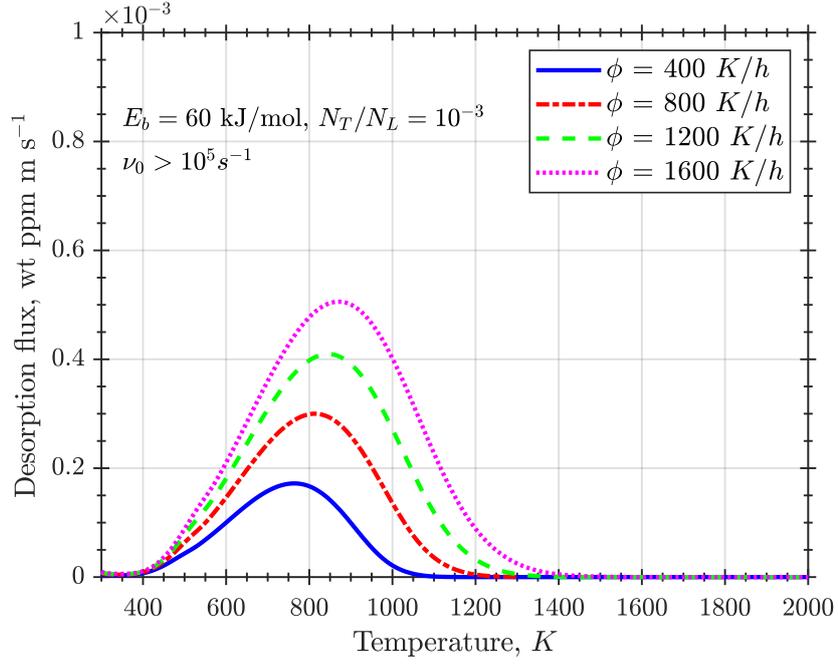

*(a)*

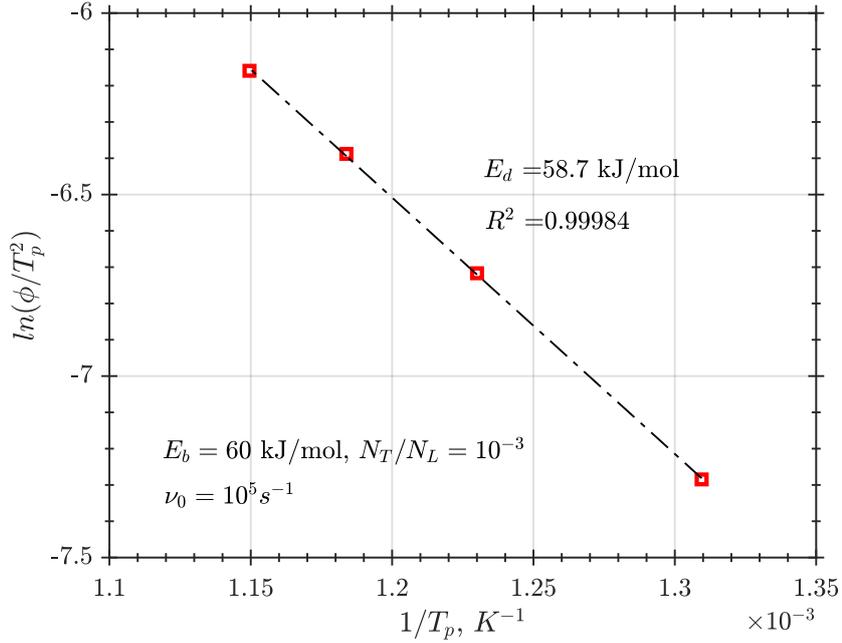

*(b)*

*Figure 6. Results for and high vibration frequency ($\nu_0 = 10^5\ s^{-1}$), strong traps ($E_b = 60$ kJ/mol) and high trap densities ($N_T = 10^{-3} N_L$). (a) Desorption flux at different heating rates; (b) detrapping energy determination using peak temperatures at different heating rates.*

## 5. Influence of charging conditions
### 5.1. Initial concentration

Numerical modelling of Thermal Desorption Analysis traditionally assumes an initial uniform lattice concentration. This follows the expectation that, after enough hydrogen charging (gaseous or electrolytic), the lattice hydrogen concentration (that only depends on solubility and fugacity) is uniform throughout the specimen. However, the initial occupancy of traps must be defined in order to solve the transport problem for both equilibrium or kinetic assumptions. After charging, it also must be assumed that (steady-state) equilibrium conditions are fulfilled so that the occupancy of hydrogen traps $\theta_T$ is univocally determined through equation (8) from the initial lattice concentration, $C_{L,0}$; the density of lattice sites, $N_L$; the binding energy of the evaluated traps, $E_b$; and the temperature during charging, $T$.

Whether traps are saturated ($\theta_T \approx 1$) or not ($\theta_T < 1$) at the beginning of heating is crucial for the balance trapping – detrapping given by McNabb and Foster's formulation. To understand the saturation regimes for different binding energies and trap densities, the relationship between $C_L$ and $C_T$ is plotted in Figure 7 assuming room temperature. It can be seen that trap density only influences the amount of trapped hydrogen, but not saturation. Both strong and weak traps are full for input lattice concentration $C_{L,0} > 1.0$ wt ppm; however, the order of magnitude of $C_T$ highly depends on binding energy for low lattice concentrations.

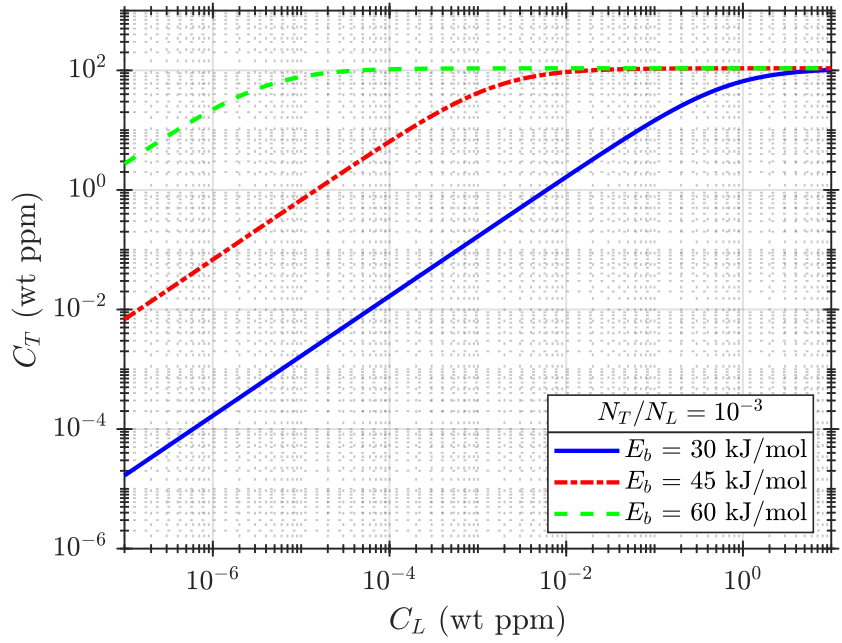

*(a)*

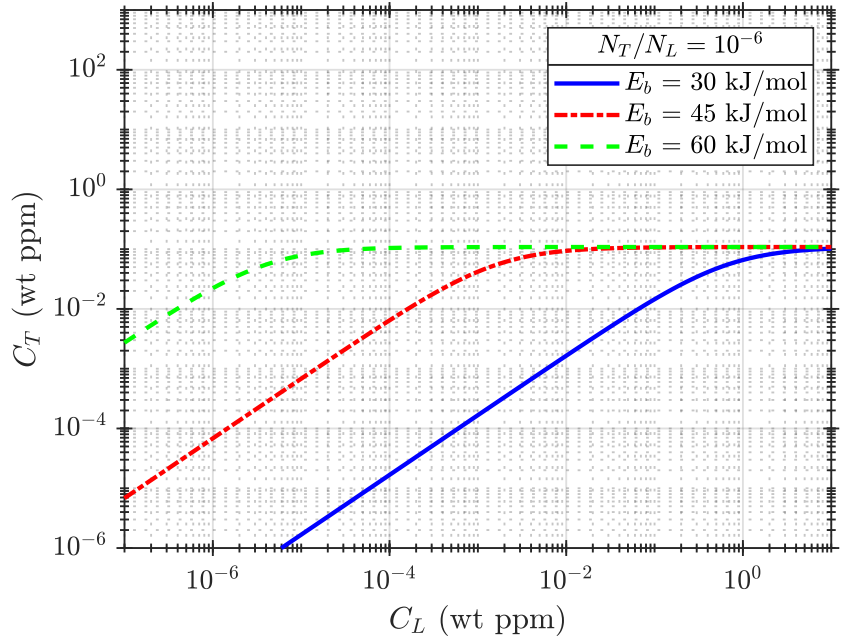

*(b)*

*Figure 7. Relationship between hydrogen lattice concentration and hydrogen trapping concentration at room temperature when thermodynamic equilibrium is assumed considering different binding energies and (a) $N_T = 10^{-3} N_L$, (b) $N_T = 10^{-6} N_L$.*

In order to assess the possible influence of $C_{L,0}$ on TDA spectra, as done before in a previous study on hydrogen permeation modelling [41], a weak trapping phenomenon ($E_b$ = 30 kJ/mol) is simulated; for low binding energy and room temperature uniform charging, the segregation regime plotted in Figure 7 hints a possible high influence of initial concentration $C_{L,0}$. Thus, due to the wider range of lattice concentration for which saturation is not achieved, only weak traps are here considered.

As shown in Figures 8 and 9, desorption peaks occur slightly earlier for $C_{L,0}$ = 1.0 wt ppm in comparison to $C_{L,0}$ = 10$^{-3}$ wt ppm. The flux magnitude, as obviously expected, is much higher for the high initial concentration case; the curves obtained for each concentration are plotted together in Figure 10. Despite the small shift in desorption spectra considering different $C_{L,0}$, when the detrapping energy is fitted through the analytic approach, it is found that the underestimation of the binding energy is less critical for $C_{L,0}$ = 10$^{-3}$ wt ppm (Figure 8.b), in which $E_b$ = 34.9 – 8.49 = 26.4 kJ/mol, than for $C_{L,0}$ = 1.0 wt ppm (Figure 9.b), in which $E_b$ = 28.9 – 8.49 = 20.4 kJ/mol.

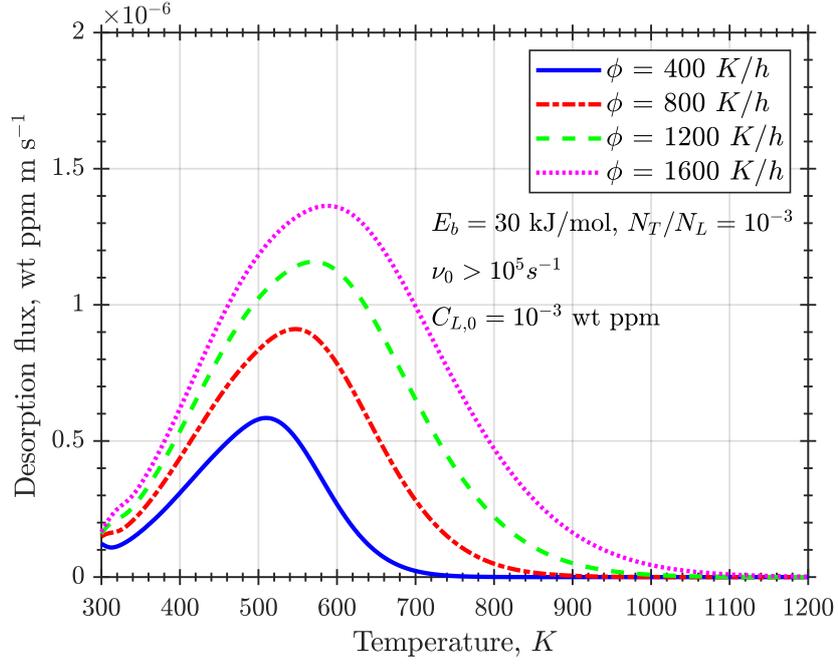

*(a)*

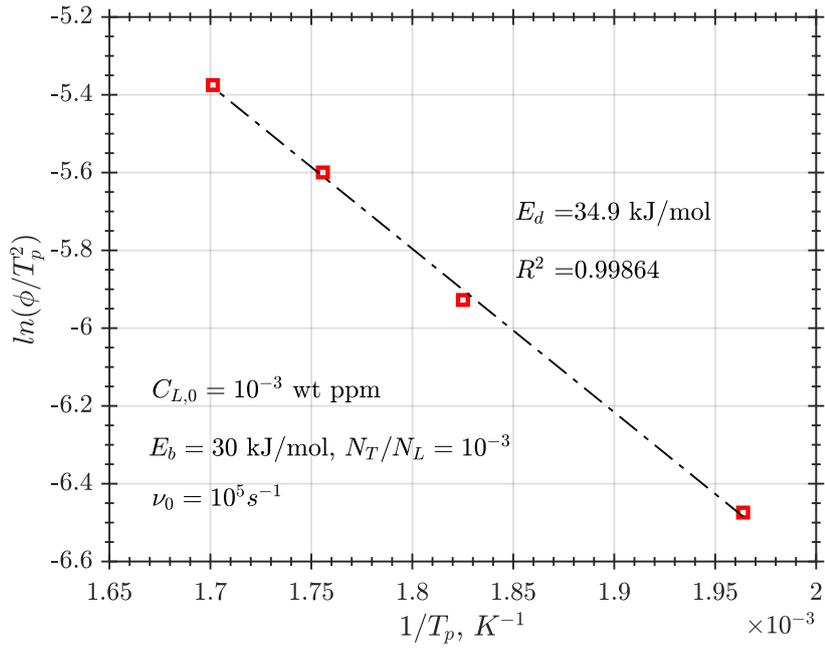

*(b)*

*Figure 8. Results for and low initial concentration ($C_{L,0}$ = 10⁻³ wt ppm), weak traps ($E_b$ = 30 kJ/mol) and high trap densities ($N_T = 10^{-3} N_L$). (a) Desorption flux at different heating rates; (b) detrapping energy determination using peak temperatures at different heating rates.*

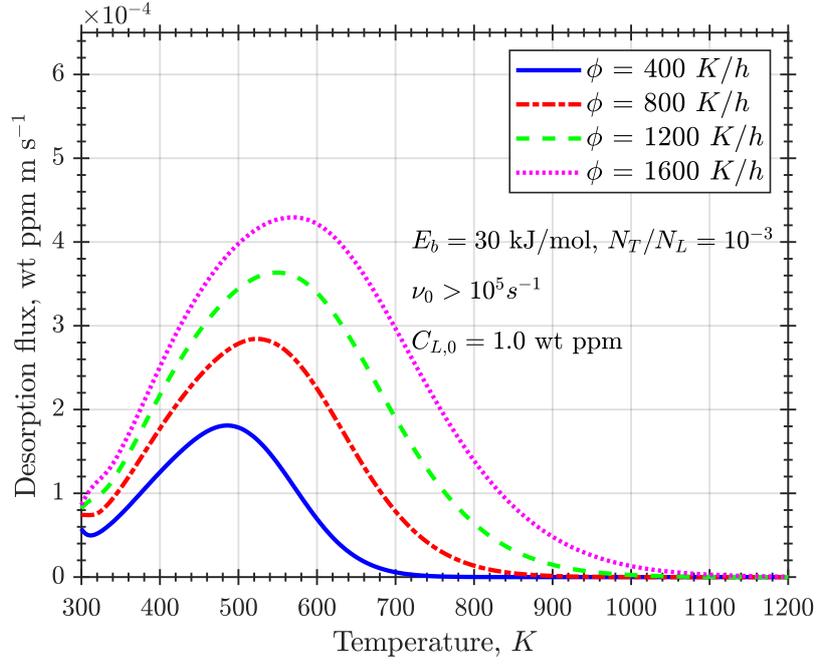

*(a)*

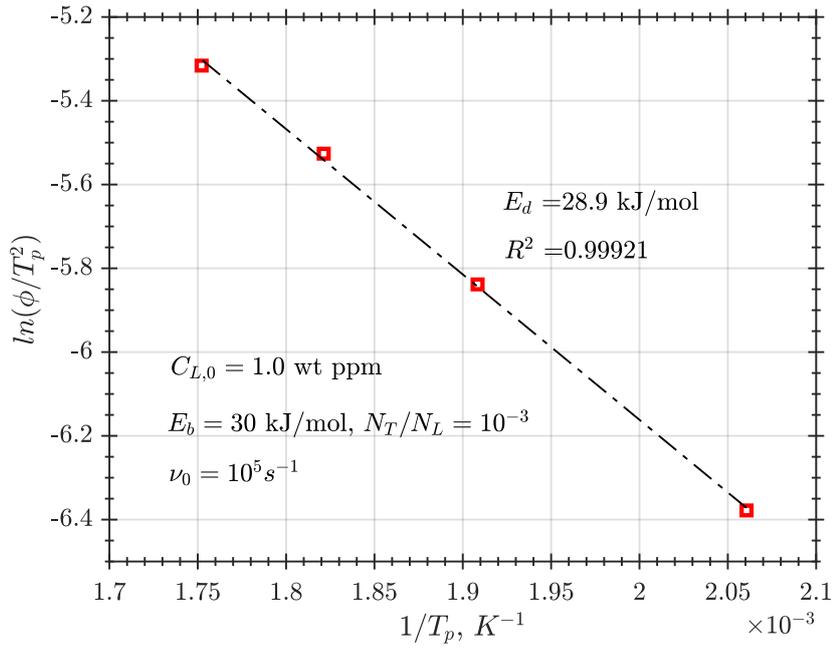

*(b)*

*Figure 9. Results for and low initial concentration ($C_{L,0}$ = 1.0 wt ppm), weak traps ($E_b$ = 30 kJ/mol) and high trap densities ($N_T = 10^{-3} N_L$). (a) Desorption flux at different heating rates; (b) detrapping energy determination using peak temperatures at different heating rates.*

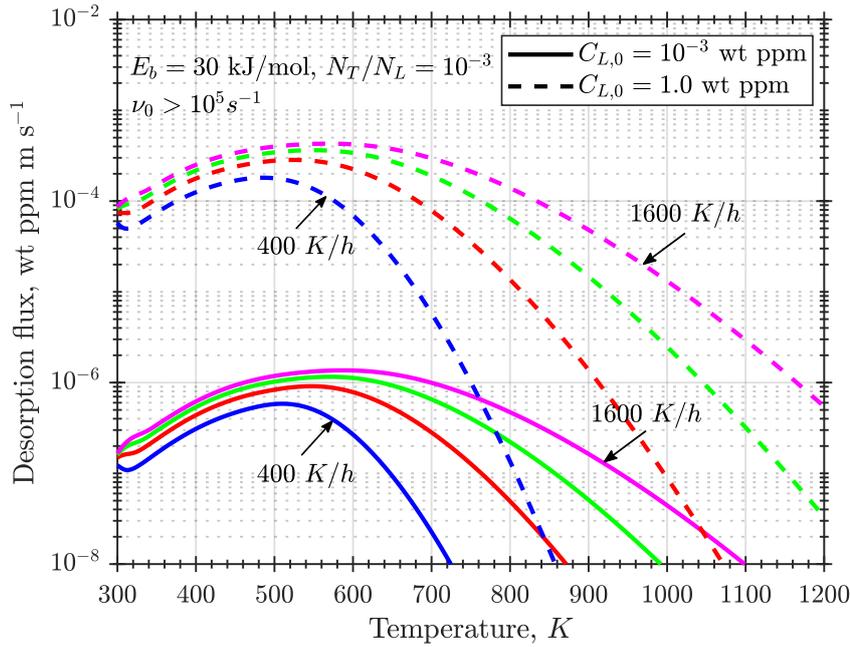

*Figure 10. Influence of initial concentration on desorption flux during programmed temperature ramp at different heating rates.*

### 5.2. Gaseous charging at high temperature and cooling influence

Section 5.1. considers a uniform charging at room temperature; however, gaseous charging is often performed at high temperatures to reach solubility values that can show embrittlement effects. Silverstein *et al.* [42] compared cathodic (50 mA·cm$^{-2}$ in 0.5 N $H_2SO_4$) and gaseous charging (300ºC and 60 MPa) effects on TDA results in lean duplex steels. Very different values of detrapping energies were found for each charging method, which was attributed to different microstructural changes induced by hydrogen: a strain-induced transformation and a damaged surface were assumed for cathodic charging while the gaseous hydrogen supposedly enhanced an intermetallic compound phase. However, in the absence of hydrogen-induced microstructural changes, charging conditions should not influence trap characterisation. Additionally, if two charging methods have equivalent fugacities it should be hypothesised that the same detrapping energies will be obtained. However, in gaseous charging, a high temperature is kept during charging in order to attain concentrations similar to those relevant to electrochemical charging. The time elapsed between the end of high-temperature charging and the beginning of a temperature-programmed ramp, i.e. sample cooling and transport from the $H_2$ high-pressure chamber to the TDA equipment, is likely to influence hydrogen distribution in such a way that desorption peaks are different to those that would be found with the homogeneous distribution. Some works have simulated all the experimental steps: (i) charging, (ii) an ageing time after charging at room temperature, and (iii) programmed-temperature desorption [29,43]. Liu et al. have studied the equivalences between electrolytic and gaseous charging and they have also analysed the implications in a TDA test [44]. In the present subsection, three steps are simulated: charging at high temperature, cooling and programmed desorption. In the following subsection, an ageing stage is included before TDA. This process is represented through the temperature evolution against time in Figure 11.

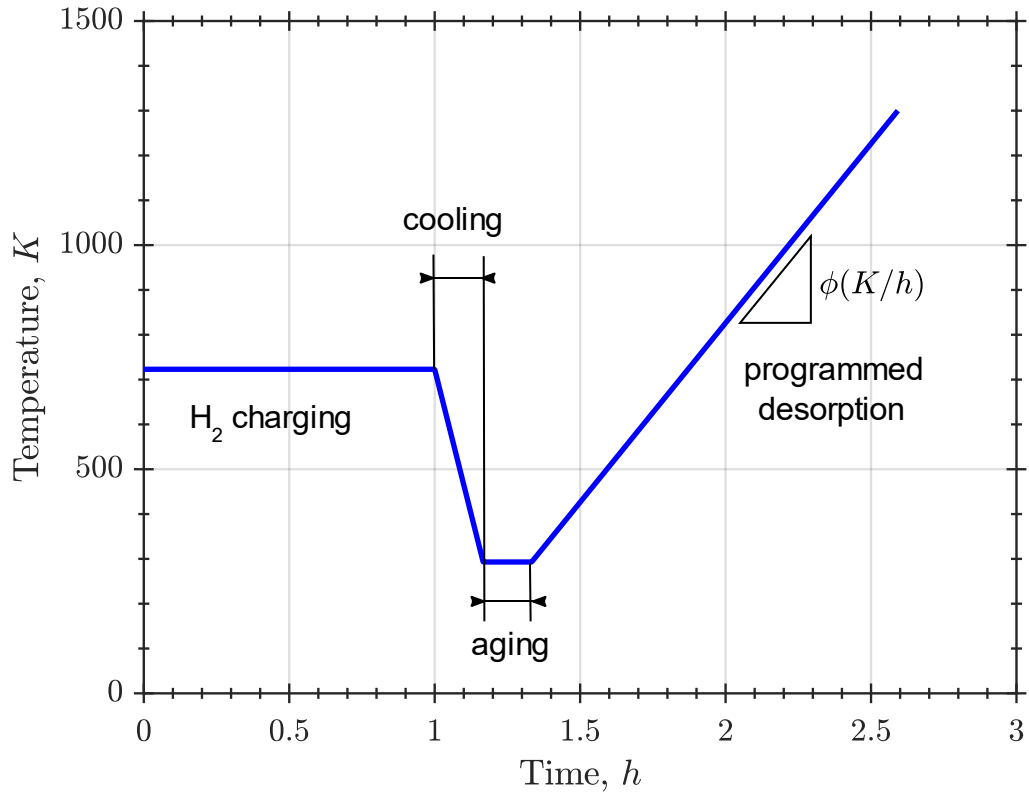

*Figure 11. Temperature evolution during hydrogen charging at high temperature, cooling, aging at room temperature (20 ºC) and thermally programmed desorption.*

Temperature charging affects hydrogen entry since the solubility is temperature-dependent. Assuming low charging pressures, hydrogen fugacity might be assumed as the gaseous $H_2$ pressure, $p_{H_2}$.

$$C_{L,0} = K_0 \exp\left(-\frac{E_s}{RT}\right)\sqrt{p_{H_2}} \qquad (9)$$

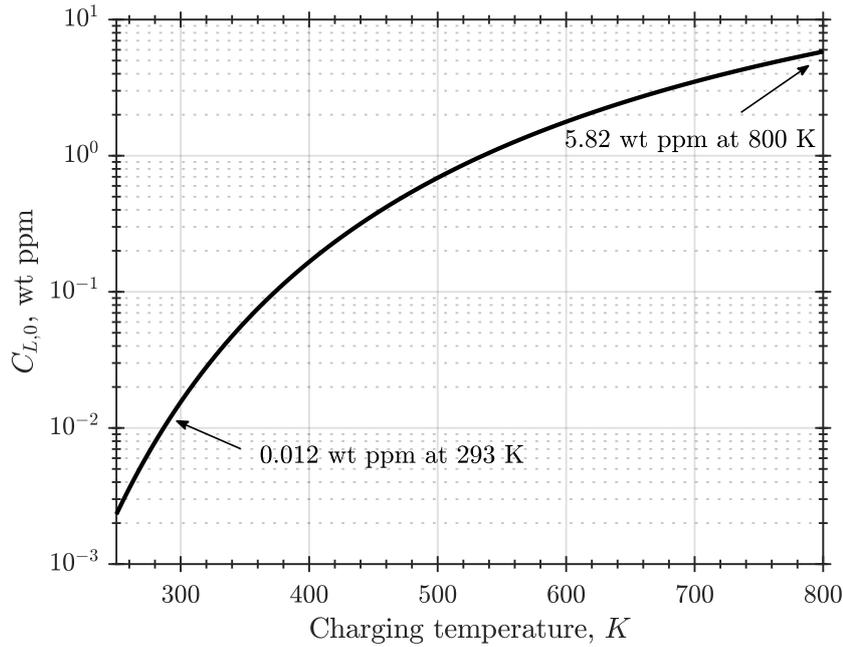

*Figure 12. Relationship between input concentration and charging temperature at 20 MPa of gaseous $H_2$ and considering $K_0$ = 45.6 wt ppm·$MPa^{0.5}$ and $E_s$ = 23.66 kJ/mol.*

For pure iron, $K_0$ is assumed to be 45.6 wt ppm·$MPa^{0.5}$ and $E_s$ equals 23.66 kJ/mol [45]. The exponential Arrhenius behaviour explains the huge difference between room temperature and high-temperature charging. As shown in Figure 12, for the considered temperature-dependent solubility and a gaseous pressure of 20 MPa, $C_{L,0}$ ranges from 0.012 wt ppm at room temperature (20ºC) to 5.82 wt ppm at 800 K. To avoid misunderstanding, it is worth mentioning that at high temperature the equilibrium changes so the relationship between $C_{L,0}$ and $C_{T,0}$ no longer follows the regimes shown in Figure 7. Segregation is plotted for high trap density and 300ºC, i.e. 573 K, in Figure 13; comparing this behaviour to that at room temperature with the same trap density (Figure 7.a) it is found that saturation of traps is shifted to higher $C_{L,0}$ concentrations.

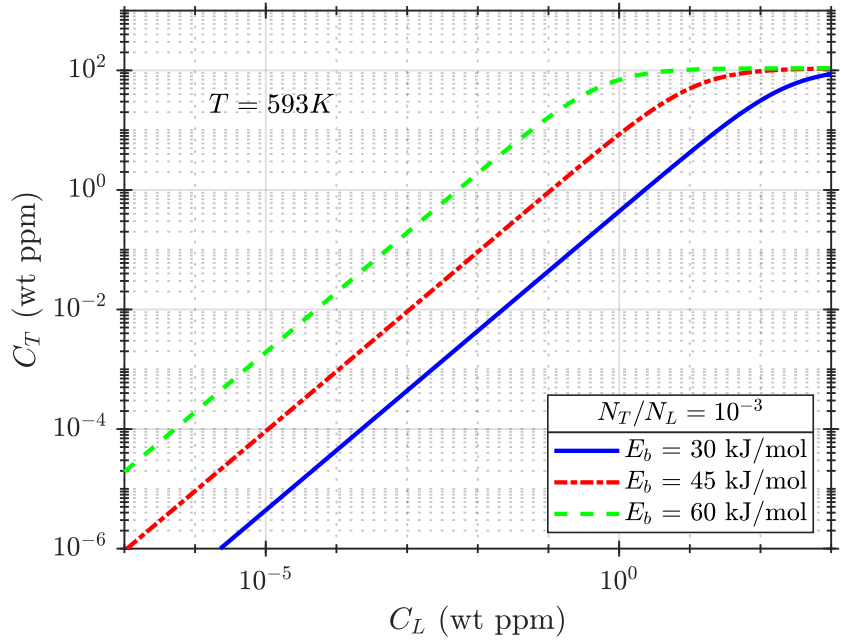

*Figure 13. Relationship between hydrogen lattice concentration and hydrogen trapping concentration at room temperature when thermodynamic equilibrium is assumed considering different binding energies and (a) $N_T = 10^{-3} N_L$, (b) $N_T = 10^{-6} N_L$.*

Gaseous hydrogen is simulated for a charging time of 5 hours at different charging temperatures. First, a fixed cooling rate is imposed; the slope of temperature evolution at the cooling stage, as depicted in Figure 11, is taken as -400 K/h. Charging influence is analysed by simulating different charging temperatures, which has a two-fold influence: (i) hydrogen input concentration $C_{L,0}$ is higher due to the temperature-dependent solubility, as shown in Figure 12; (ii) cooling takes more time for higher charging temperatures because the cooling rate is fixed. It must be recalled that hydrogen charging ends when the cooling step begins, i.e. during cooling a zero-concentration is imposed as boundary condition so desorption starts. Competition between both effects must be evaluated for each specific situation: high charging temperatures $T_{ch}$ enhance hydrogen solubility but imply a longer cooling stage; thus, they can diminish the initial concentration at the beginning of TDA test. This contradictory result is found for different charging temperatures between 293 and 693 K when a low binding energy is considered ($E_b$ = 30 kJ/mol), as shown in Figure 14.

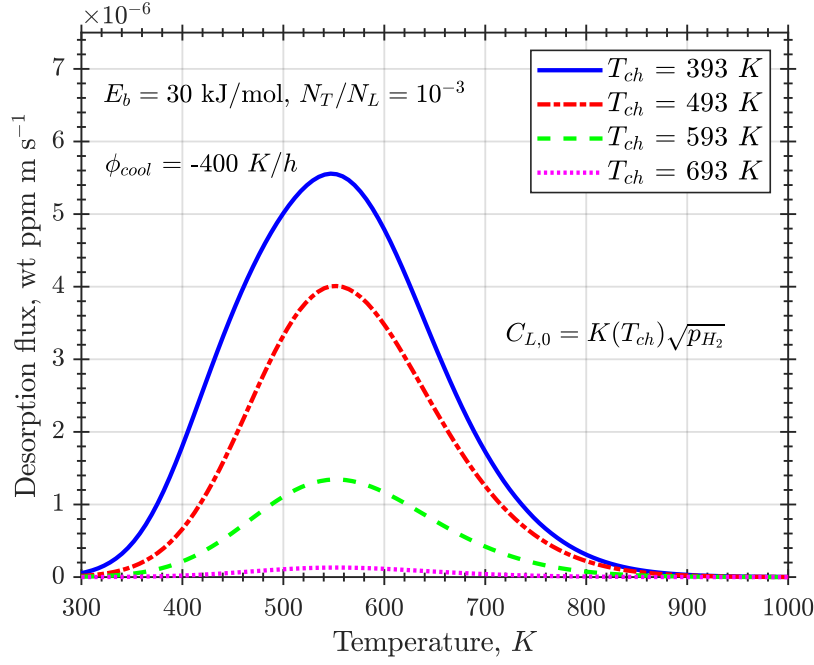

Figure 14. Influence of charging temperature on desorption flux for a charging time of 5 hours and a fixed cooling rate of -400 K/h considering weak traps ($E_b$ = 30 kJ/mol).

The evolution of the integrated lattice concentration over the specimen, $\langle C_L(t) \rangle = \frac{1}{a}\int_0^a C_L(x,t)dx$, is plotted in Figure 15 to analyse the counterintuitive fact that higher charging temperatures produce smaller desorption curves. As expected, hydrogen rapidly fills lattice sites so saturation $C_{L,0} = K(T_{ch})\sqrt{20\ MPa}$ is attained. However, due to the longer cooling stage and the high diffusivity of hydrogen at 693 K, desorption during cooling is critical, reducing thus drastically the amount of hydrogen in interstices. A detail of this evolution of integrated $\langle C_L(t) \rangle$ is shown in Figure 15.b. This detail is intended to show the end of the cooling stage for each $T_{ch}$. It can be clearly seen how the flux rises when cooling ends and the temperature ramp begins. Representing the evolution of integrated trapping concentration, $\langle C_T(t) \rangle = \frac{1}{a}\int_0^a C_T(x,t)dx$ in Figure 16, the segregation behaviour that has been previously represented in Figure 7 for room temperature and in Figure 13 for 593 K plays an important role. The small peaks of trapped hydrogen during desorption in Figure 16 show that not only detrapping is happening, but also retrapping phenomena. For high temperatures, the equilibrium concentration in trapping sites $C_T$ is lower even though $C_{L,0}$ is higher. Actually, at room temperature the trapping process is slower because trapping is also is temperature-dependent following an exponential function, as expressed in equation (7), so the charging time of 5 hours is not enough to reach the high asymptotic equilibrium $C_T$. During the cooling stage detrapping slows down, while trapping is still active, and both effects induce an increase of $\langle C_T(t) \rangle$. Thus, a peak of trapped hydrogen $\langle C_T(t) \rangle$ is observed in Figure 16 for weak traps and a plateau is found in Figure 19 because hydrogen is trapped deeper.

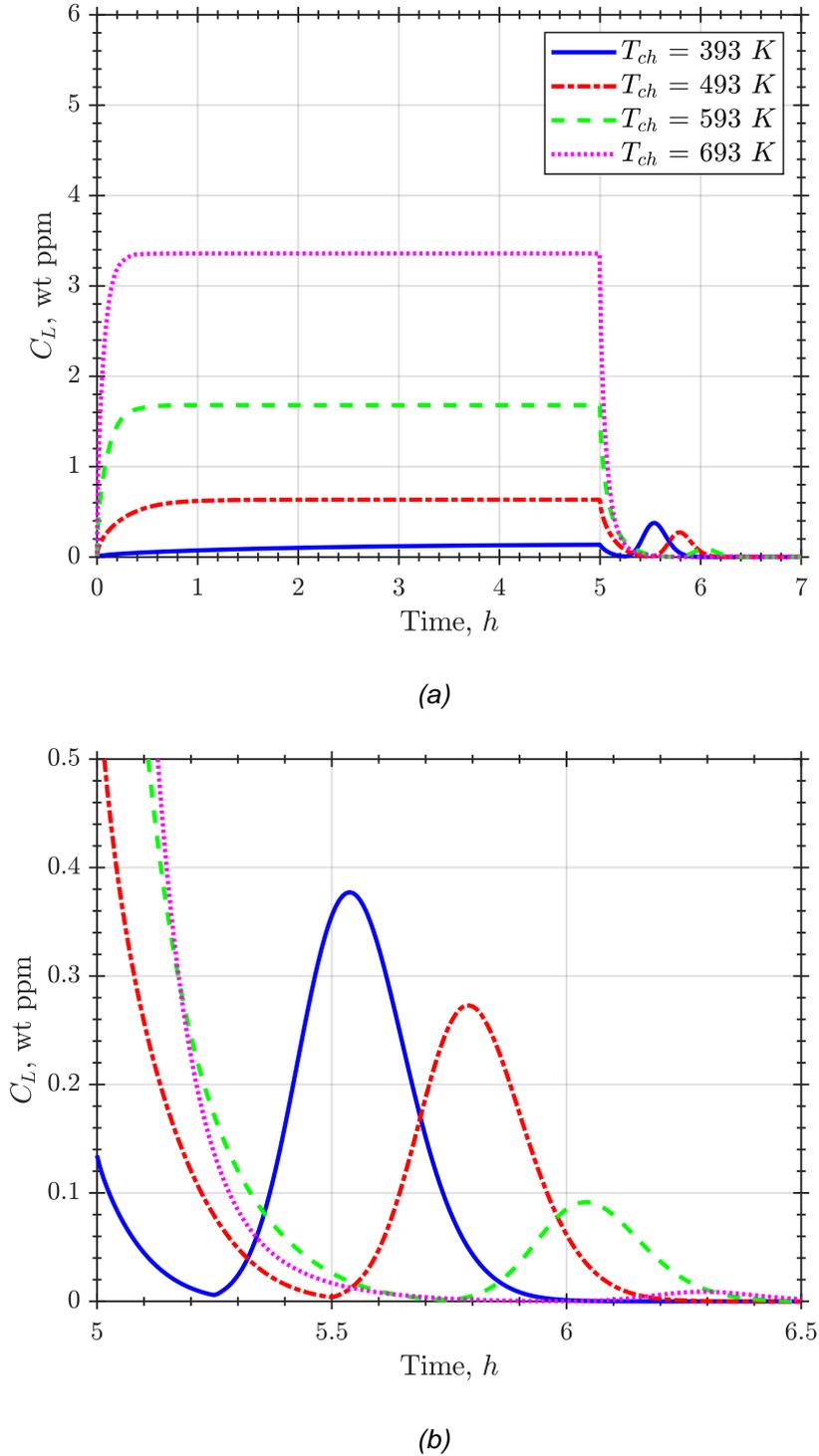

*Figure 15. Influence of charging temperature on the evolution of lattice hydrogen ($C_L$) for a charging time of 5 hours and a fixed cooling rate of -400 K/h considering weak traps ($E_b$= 30 kJ/mol). (a) Evolution during the whole process; (b) detail of $C_L$ evolution during cooling and TDA.*

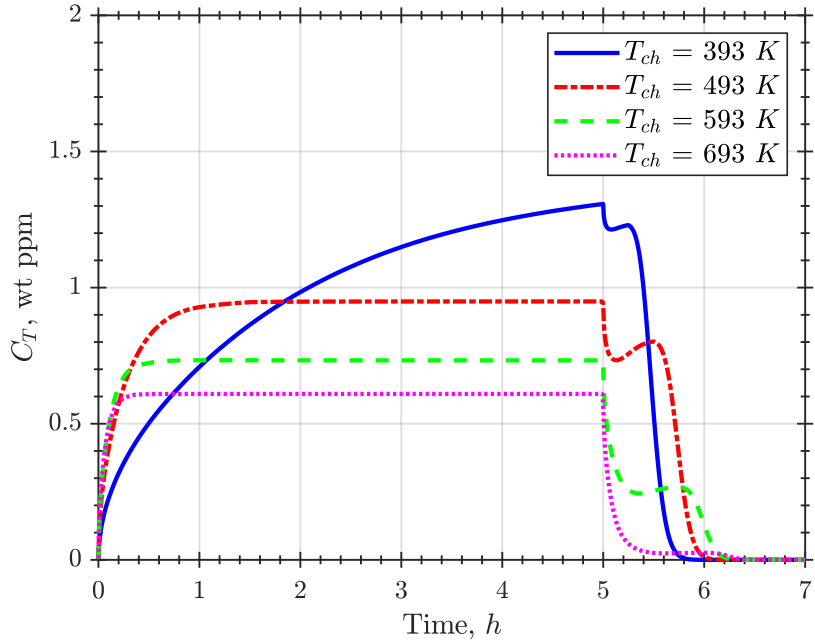

*Figure 16. Influence of charging temperature on the evolution of trapped hydrogen ($C_T$) for a charging time of 5 hours and a fixed cooling rate of -400 K/h considering weak traps ($E_b$ = 30 kJ/mol).*

However, this behaviour changes for strong traps ($E_b$ = 60 kJ/mol), as shown in Figure 17 (TDA spectra), Figure 18 ($\langle C_L(t) \rangle$ evolution) and Figure 19 ($\langle C_T(t) \rangle$ evolution). Important differences in comparison to weak traps are observed; the maximum amount of diffusible hydrogen, i.e. $\langle C_L(t) \rangle_{max}$, is very high for charging temperatures of 593 and 693 K. For all cases $\langle C_L(t) \rangle_{max}$ is higher than the charging equilibrium $C_{L,0}$. Additionally, the order of magnitude of $C_T$ (10 – 100 wt ppm) is above the usual experimental values. This is due to the high binding energy and high trap density. However, after desorption during cooling, the order of magnitude of desorption flux during TDA is more realistic. As shown in Figure 19, only for $T_{ch}$ = 693 K the asymptotic equilibrium $C_T$ is reached before 5 hours. While for $T_{ch}$ = 593 K the maximum flux is higher than the corresponding curve for $T_{ch}$ = 693 K, the situation is inverted for $T_{ch}$ = 393 K versus $T_{ch}$ = 493 K. This contradictory result can be explained because at 693 K the saturation $C_T$ value is lower than the value at 593 K; however, at 693 K hydrogen moves faster so traps are completely filled after 5 hours only at a charging temperature of 693 K. Thus, more trapped hydrogen is available after 593 K than after 693 K. Obviously, for longer charging times to achieve saturation for every charge this contradictory result would not be found. These latter lower temperatures also show a previous peak at low temperatures that hides the secondary peak occurring at the same temperature than the maximum flux for 593 and 693 K. It is shown that the competition between trapping equilibrium during charging, desorption during cooling and detrapping during the TDA ramp is hard to predict, so each specific combination of charging conditions and trapping features must be evaluated.

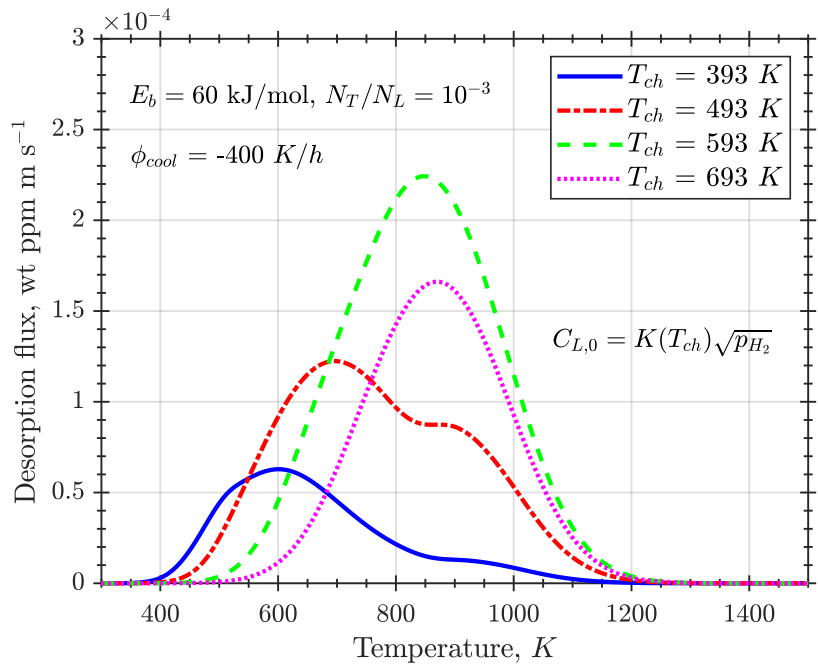

Figure 17. Influence of charging temperature on desorption flux for a charging time of 5 hours and a fixed cooling rate of -400 K/h considering strong traps ($E_b$ = 60 kJ/mol).

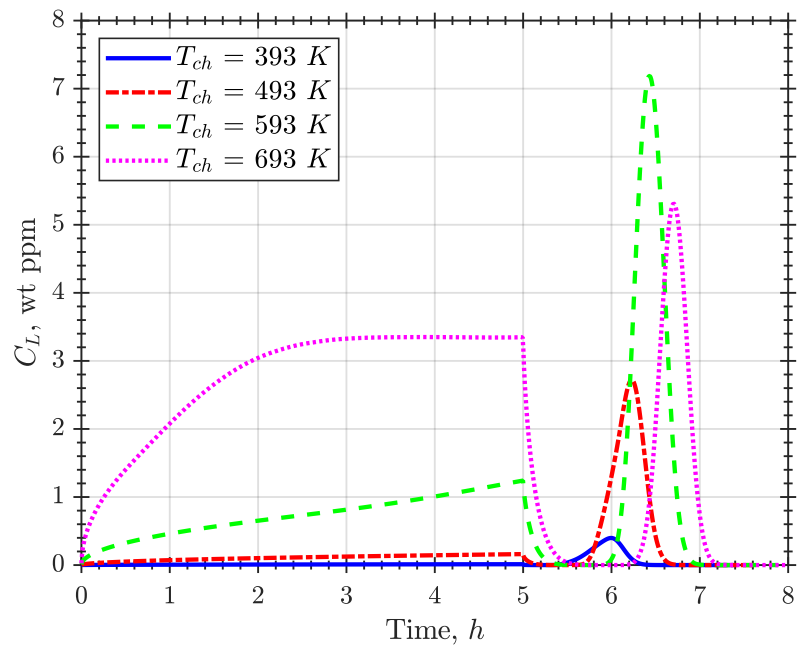

*Figure 18. Influence of charging temperature on the evolution of lattice hydrogen ($C_L$) for a charging time of 5 hours and a fixed cooling rate of -400 K/h considering strong traps ($E_b$ = 60 kJ/mol).*

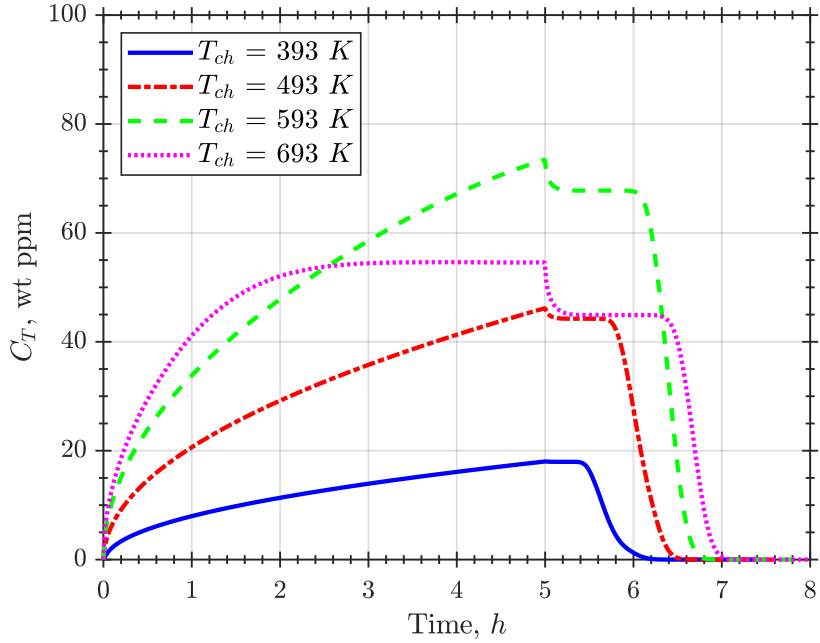

*Figure 19. Influence of charging temperature on the evolution of trapped hydrogen ($C_T$) for a charging time of 5 hours and a fixed cooling rate of -400 K/h considering strong traps ($E_b$ = 60 kJ/mol).*

With the aim of simplifying the analysis of detrapping processes, the cooling time effect, i.e. the longer time required for high charging temperatures, is decoupled by fixing a cooling time of 1 hour instead of a temperature ramp. In this case, as shown in Figure 21, the maximum value of $\langle C_L(t) \rangle$ occurs at the same time, independently of the charging temperature. However, as might be seen in Figures 20 to 25, this assumption does not change significantly the TDA spectra and the effects of gaseous charging and trapping parameters on desorption flux. Thus, the same conclusions drawn for the fixed $\phi_{cool}$ = -400 K are valid for a controlled cooling time, $t_{cool}$ = 1 h, from charging temperature to 20ºC.

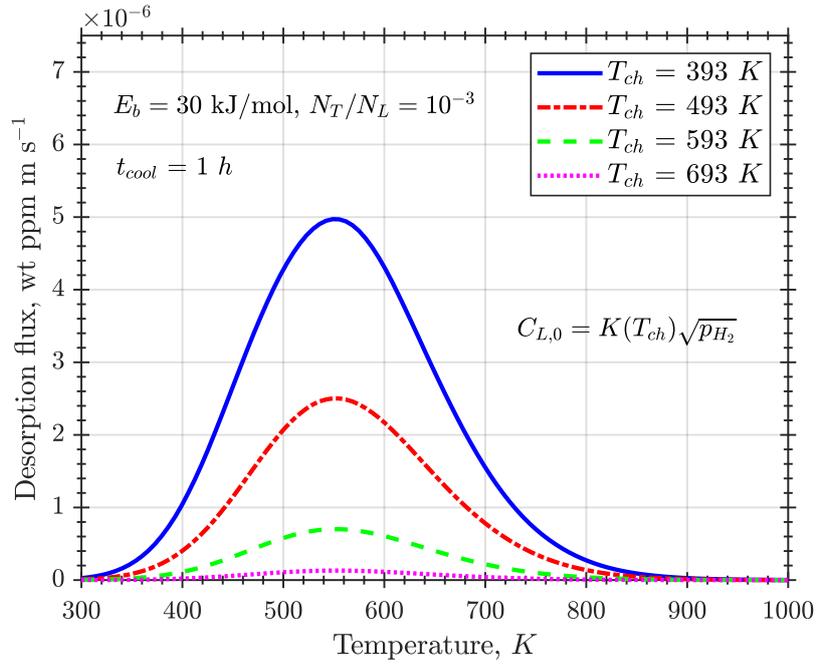

*Figure 20. Influence of charging temperature on desorption flux for a charging time of 5 hours and a fixed cooling time of 1 hour considering weak traps ($E_b$ = 30 kJ/mol).*

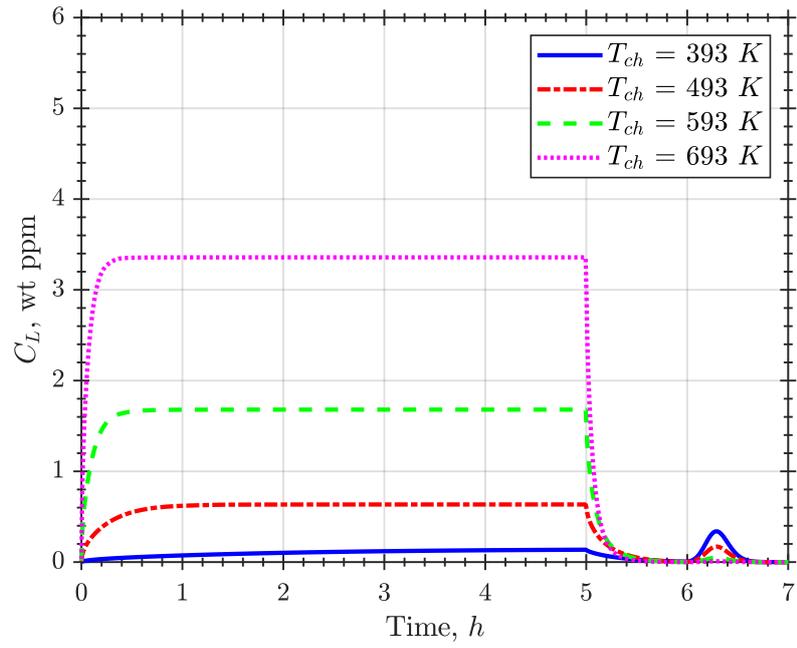

(a)

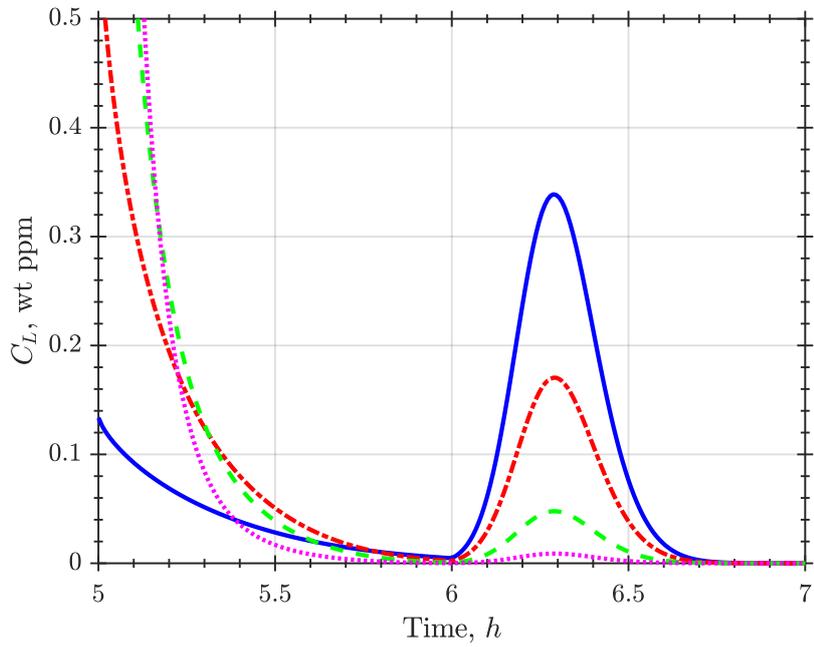

(b)

*Figure 21. Influence of charging temperature on the evolution of lattice hydrogen ($C_L$) for a charging time of 5 hours and a fixed cooling time of 1 hour considering weak traps ($E_b$ = 30 kJ/mol). (a) Evolution during the whole process; (b) detail of $C_L$ evolution during cooling and TDA.*

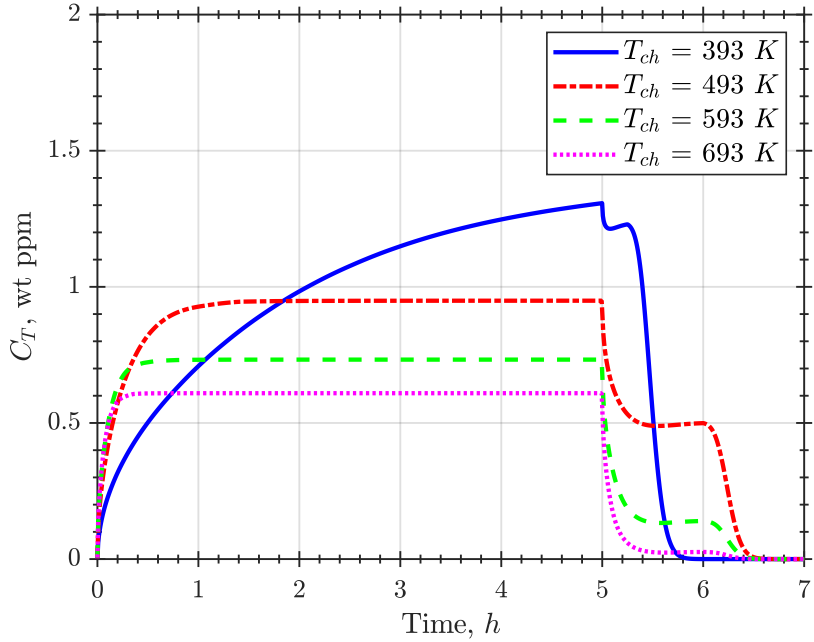

*Figure 22. Influence of charging temperature on the evolution of trapped hydrogen ($C_T$) for a charging time of 5 hours and a fixed cooling time of 1 hour considering weak traps ($E_b$= 30 kJ/mol).*

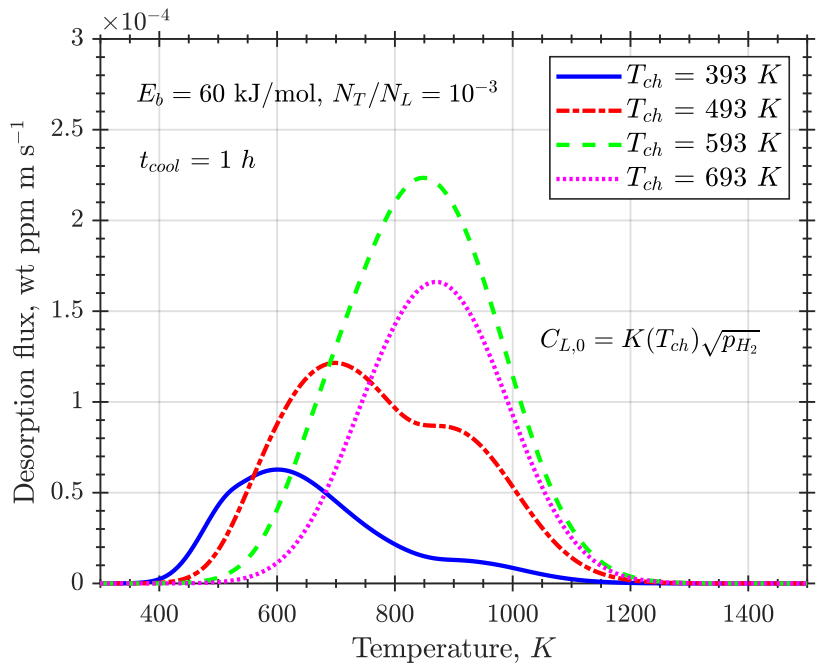

*Figure 23. Influence of charging temperature on desorption flux for a charging time of 5 hours and a fixed cooling time of 1 hour considering strong traps ($E_b$= 60 kJ/mol).*

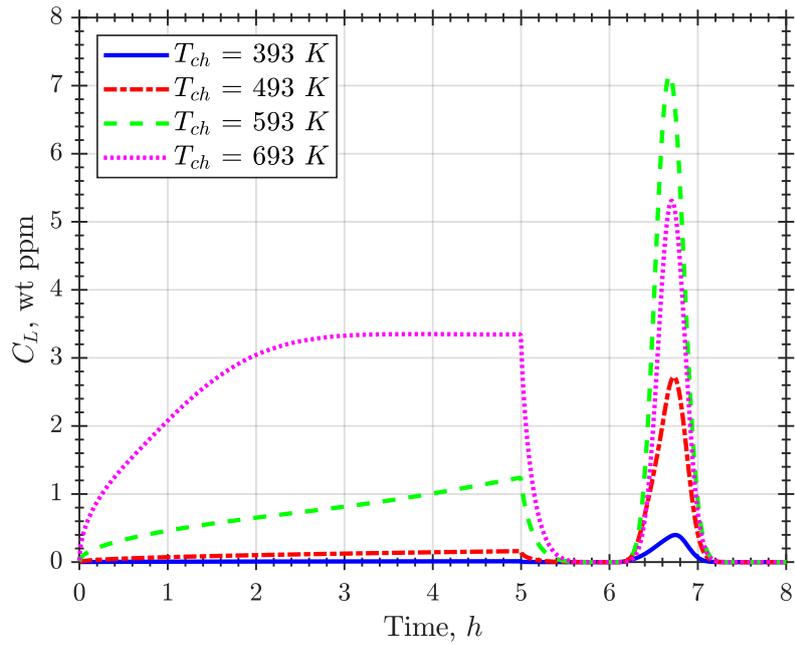

*Figure 24. Influence of charging temperature on the evolution of lattice hydrogen ($C_L$) for a charging time of 5 hours and a fixed cooling time of 1 hour considering strong traps ($E_b$ = 60 kJ/mol).*

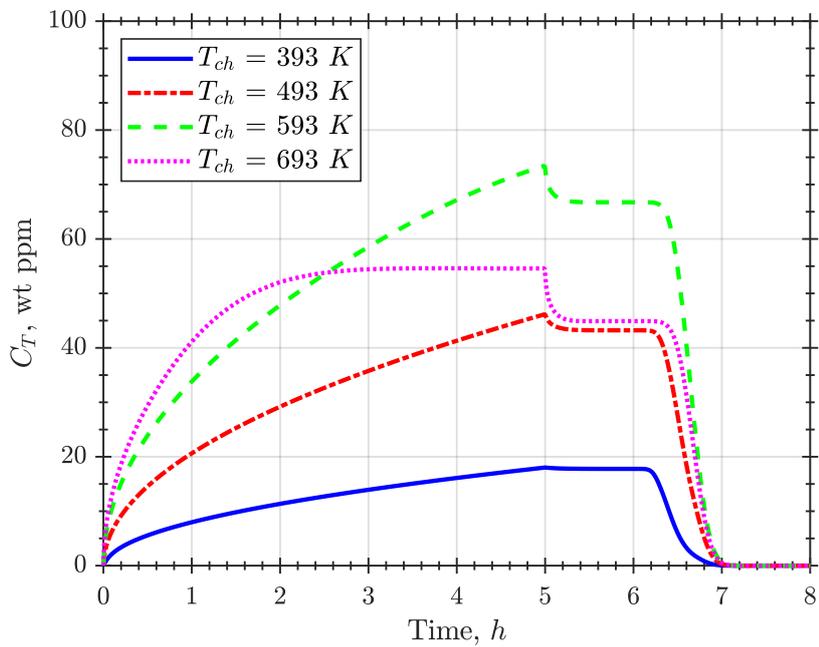

*Figure 25. Influence of charging temperature on the evolution of trapped hydrogen ($C_T$) for a charging time of 5 hours and a fixed cooling time of 1 hour considering strong traps ($E_b$ = 60 kJ/mol).*

Finally, the influence of the charging time is studied for a high temperature (593 K) since it has been previously observed that $C_T$ saturation is not reached at 5 hours. As shown in Figure 26, for higher charging times, the desorption flux curve is slightly bigger but there is no difference between results corresponding to 7.5 and 10 hours because traps are completely saturated at a time between 5 and 7.5 hours. However, for a lower $t_{ch}$ of 2.5 hours, the TDA flux evolution is not just lower but a secondary peak at low temperatures appears, as observed for $t_{ch}$ = 5 h but lower charging temperatures. This secondary peak is occurring due to a similar fact that can be shown in Figures 17, 18 and 19 and it is attributed to retrapping processes. The emptier that strong traps are after charging, the more retrapping effects will be operative during TDA.

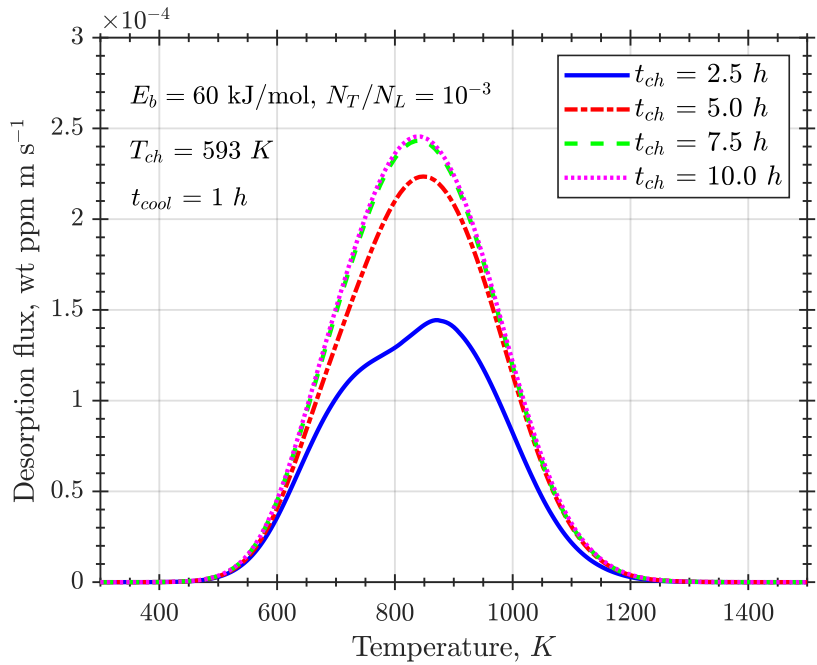

*Figure 26. Influence of charging time on desorption flux for a charging temperature of 593 K and a fixed cooling time of 1 hour considering strong traps ($E_b$ = 60 kJ/mol).*

### 5.3. Ageing influence

In addition to the required cooling step after gaseous charging, which cannot be avoided for TDA starting from room temperature, a resting step may come into play. This is due to the need for transporting the sample, which adds an elapsed time at room temperature before the temperature ramp begins.

Raina et al. [40] studied the influence of rest time at room temperature. However, in the present work, this time is considered after a high-temperature gaseous charging and a cooling step, so hydrogen distribution is not uniform at the beginning of ageing.

For strong traps (60 kJ/mol), ageing at room temperature has no impact on TDA results, as shown in Figure 27. This is because, for high-temperature gaseous charging, lattice sites are completely depleted before ageing starts; while detrapping during ageing is not possible due to the strong retention of hydrogen in traps. However, for weak traps, detrapping is progressively happening during ageing, as might be verified in Figure 28.b so the final observed desorption flux during a programmed temperature ramp is lower for long ageing times (Figure 28.a).

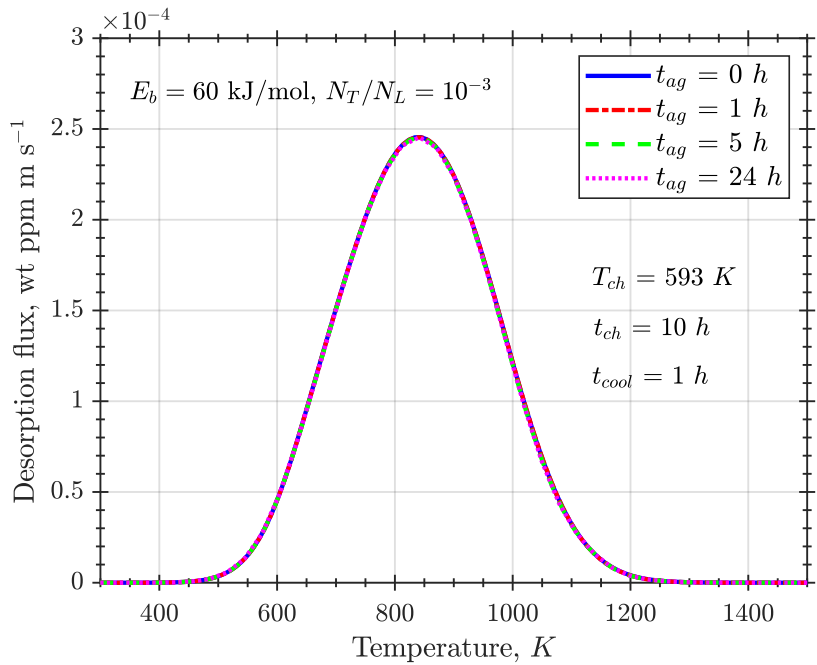

Figure 27. Influence of aging time on desorption flux for a charging temperature of 593 K and a fixed cooling time of 1 hour considering strong traps ($E_b$ = 60 kJ/mol).

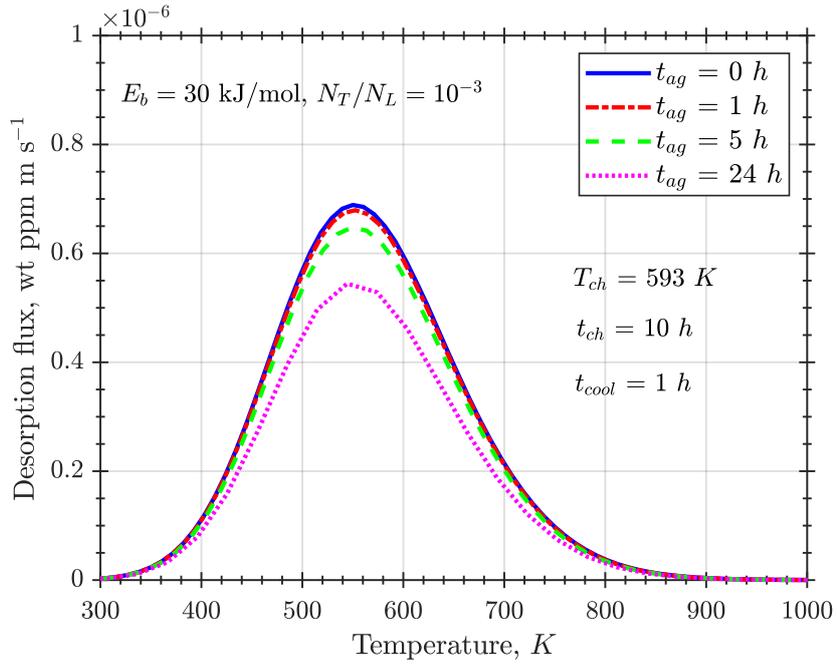

*(a)*

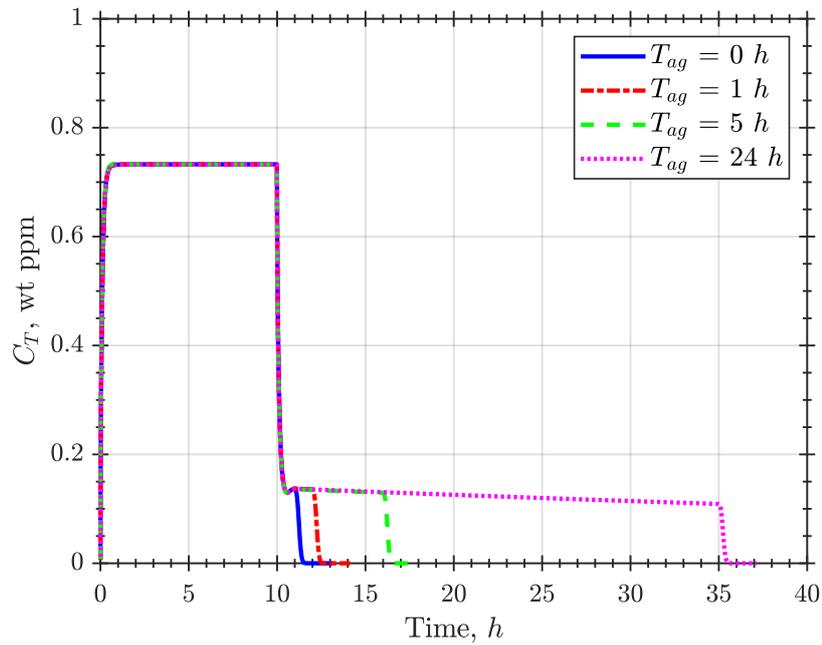

*(b)*

*Figure 28. Influence of aging time on (a) desorption flux and (b) evolution of trapped hydrogen ($C_T$), for a charging temperature of 593 K and a fixed cooling time of 1 hour considering weak traps ($E_b$ = 30 kJ/mol).*

## 6. Conclusions

Modelling hydrogen transport within metals is essential to understand the coupled diffusion – damage phenomena operating during embrittlement. To characterise trapping effects, thermally programmed desorption is the most common technique due to its simplicity for obtaining detrapping energies. However, this experimental technique, here referred to as Thermal Desorption Analysis (TDA), does not have spatial resolution and trapping features are derived from the appearance of a desorption maximum that is related to detrapping processes. The generic formulation first proposed by McNabb and Foster must be used to model these complexities, as the commonly used analytic approach based on Kissinger's expression is too simplistic. In the present paper, charging conditions for gaseous hydrogen entry at high temperatures are investigated within the McNabb-Foster modelling context with the objective of achieving a better understanding of TDA spectra. The governing equation of hydrogen diffusion is solved in a finite element code in which the generic kinetic expression modelling trapping – detrapping has been also included. First, the equilibrium validity of Oriani has been verified for different vibration frequencies. It has been concluded that frequency effects are only visible for very low values (< $10^5$ s$^{-1}$); thus, frequencies of the order of Debye frequency ($10^{13}$ s$^{-1}$) produce a behaviour, for the heating rates and diffusivity values here considered, completely equivalent to thermodynamic equilibrium between hydrogen in lattice and trapping sites. This observation is verified for both weak and strong traps.

Charging conditions have been assessed, as a first approximation, by simulating different values of uniform $C_{L,0}$. The expected behaviour can be predicted just by plotting segregation, i.e. the relationship between $C_L$ and $C_T$ at equilibrium. For weak traps ($E_b$ = 30 kJ/mol) and high trap density, the maximum flux is much lower for $C_{L,0}$ = 10$^{-3}$ wt ppm in comparison to $C_{L,0}$ = 1.0 wt ppm, as expected. However, for $C_{L,0}$ = 10$^{-3}$ wt ppm the desorption maximum is slightly delayed. Detrapping energies have been determined through the analytic linear model and it has been demonstrated that Kissinger's expression underestimates binding energies: simulated curves with $E_b$ = 30 kJ/mol are fitted through this procedure to $E_b$ = 26.4 kJ/mol and $E_b$ = 20.4 kJ/mol for $C_{L,0}$ = 1.0 wt ppm and for $C_{L,0}$ = 10$^{-3}$ wt ppm, respectively.

The main difference when modelling gaseous charging is related to the high temperature process in which solubility increases exponentially; here, the boundary condition during the charging step is considered for different charging temperatures and an H$_2$ pressure of 20 MPa. Depending on the trapping features, it has been shown that charging temperature and the associated cooling step affect TDA spectra. For low charging temperatures, secondary peaks have been obtained even though only one type of trap is considered in the mass balance. Ageing time is only critical for weak traps in which detrapping process occurs even at room temperature.

Simulation of charging, cooling and ageing steps reproduce a more realistic TDA test since hydrogen desorption is almost zero at the beginning of the thermally programmed ramp. This observation is explained because diffusible hydrogen escapes during cooling as this step is usually performed in the absence of hydrogen pressure. It has been demonstrated that the consideration of a uniform lattice hydrogen concentration as initial condition does not realistically reproduce TDA tests. However, other complex phenomena such as multi-trapping or microstructure evolution during heating must be included in future modelling frameworks. Future research should also focus on the inverse problem of trapping parameter determination with the aim of accurately characterising the binding energies of traps and, if possible, other features such as trapping densities or lattice diffusion parameters.

**Acknowledgements**

The authors gratefully acknowledge financial support from the Ministry of Economy and Competitiveness of Spain through grants MAT2014-58738 and RTI2018-096070-B-C33. E. Martínez-Pañeda additionally acknowledges financial support from Wolfson College (Cambridge) through their Junior Research Fellowship.